\documentclass[english,letter]{article}
\usepackage[T1]{fontenc}
\usepackage[latin9]{inputenc}
\usepackage{geometry}
\usepackage{color}
\usepackage{babel}
\usepackage{array}
\usepackage{refstyle}
\usepackage{url}
\usepackage{amsmath}
\usepackage{graphicx}
\usepackage[pdfusetitle,
 bookmarks=true,bookmarksnumbered=false,bookmarksopen=false,
 breaklinks=true,pdfborder={0 0 0},pdfborderstyle={},backref=false,colorlinks=true]
 {hyperref}

\makeatletter

\AtBeginDocument{\providecommand\secref[1]{\ref{sec:#1}}}
\AtBeginDocument{\providecommand\tabref[1]{\ref{tab:#1}}}
\AtBeginDocument{\providecommand\figref[1]{\ref{fig:#1}}}
\AtBeginDocument{\providecommand\Figref[1]{\ref{Fig:#1}}}
\AtBeginDocument{\providecommand\eqref[1]{\ref{eq:#1}}}
\providecommand{\tabularnewline}{\\}
\RS@ifundefined{subsecref}
  {\newref{subsec}{name = \RSsectxt}}
  {}
\RS@ifundefined{thmref}
  {\def\RSthmtxt{theorem~}\newref{thm}{name = \RSthmtxt}}
  {}
\RS@ifundefined{lemref}
  {\def\RSlemtxt{lemma~}\newref{lem}{name = \RSlemtxt}}
  {}

\usepackage{enumerate}
\usepackage{color}
\usepackage{afterpage}
\usepackage{longtable}

\usepackage[table]{xcolor}
\usepackage{relsize}

    \definecolor{cSignifOne}{rgb}{.92,1,.92}
    \definecolor{cSignifTwo}{rgb}{.78,1,.78}
    \definecolor{cSignifThree}{rgb}{.5,1,.5}
    \definecolor{cSignifThousandth}{rgb}{.6,1,0} %

\usepackage[aboveskip=1pt,labelfont=bf,labelsep=period,justification=raggedright,singlelinecheck=off,figurename=Fig.]{caption}

\hypersetup{allcolors=blue}

\newref{fig}{refcmd={\hyperref[#1]{Figure \ref{#1}}}}
\newref{tab}{refcmd={\hyperref[#1]{Table \ref{#1}}}}
\newref{eq}{refcmd={\hyperref[#1]{Eq. (\ref{#1})}}}
\newref{sec}{refcmd={\hyperref[#1]{Section \ref{#1}}}}

\newref{Fig}{refcmd={\hyperref[#1]{Figure \ref{#1}}}}
\newref{Tab}{refcmd={\hyperref[#1]{Table \ref{#1}}}}
\newref{Eq}{refcmd={\hyperref[#1]{Eq. (\ref{#1})}}}
\newref{Sec}{refcmd={\hyperref[#1]{Section \ref{#1}}}}

\usepackage{bbm} %
\usepackage{pdflscape} %
\usepackage{geometry}
\usepackage{array} %
\usepackage{rotating}

\usepackage{inglehartcolors} %
\usepackage{xspace}
\newcommand\GWPindivCorrLSCL{0.63\xspace
} 
\newcommand\GFSindivCorrLSCL{0.56\xspace
} 
\newcommand\yearRangeCLGWP{2006--2022\xspace
} 
\newcommand\yearRangeCLGWPTwentyOneTwentyTwoGFScountries{2021--2022\xspace
} 
\newcommand\yearRangeCLGWPmatchedtoWVS{2006--2022\xspace
} 
\newcommand\yearRangeLSWVSmatchedtoGFS{2001--2022\xspace
} 
\newcommand\yearRangeCLGFS{2023\xspace
} 
\newcommand\yearRangeLSGFS{2023\xspace
} 
\newcommand\yearRangeLSGWP{2007--2010\xspace
} 
\newcommand\yearRangeLSWVSmatchedtoGWP{2006--2022\xspace
} 
\newcommand\yearRangeLSGWPGFScountries{2007--2010\xspace
} 
\newcommand\GWPindivCorrLSCLGFScountries{0.62\xspace
} 

\usepackage{booktabs}

\usepackage{lineno}

\makeatother

\usepackage[style=authoryear,uniquename=false,uniquelist=false,backend=biber,style=authoryear,sorting=nyt,sortcites=ynt]{biblatex}
\begin{document}
\date{2 September 2025}

\title{Are international happiness rankings reliable?}
\author{C P Barrington-Leigh\thanks{Please use \protect\href{https://alum.mit.edu/www/cpbl/publications/Barrington-Leigh-DRAFT2025-global-differences-CLE.pdf}{the latest version of this paper online}.\protect \\
Author contact at \protect\url{https://wellbeing.research.mcgill.ca/contact}.
}}
\maketitle
\begin{abstract}

Global comparisons of wellbeing increasingly rely on survey questions
that ask respondents to evaluate their lives, most commonly in the
form of \textquotedblleft life satisfaction\textquotedblright{} and
\textquotedblleft Cantril ladder\textquotedblright{} items. These
measures underpin international rankings such as the World Happiness
Report and inform policy initiatives worldwide, yet their comparability
has not been established with contemporary global data. Using the
Gallup World Poll, Global Flourishing Study, and World Values Survey,
I show that the two question formats yield divergent distributions,
rankings, and response patterns that vary across countries and surveys,
defying simple explanations. To explore differences in respondents\textquoteright{}
cognitive interpretations, I compare regression coefficients from
the Global Flourishing Study, analyzing how each question wording
relates to life circumstances. While international rankings of wellbeing
are unstable, the scientific study of the determinants of life evaluations
appears more robust. Together, the findings underscore the need for
a renewed research agenda on critical limitations to cross-country
comparability of wellbeing.

\newpage

\end{abstract}
\tableofcontents{}

\listoffigures

\listoftables

\newpage

\section{Introduction}

The rising international interest in \emph{happiness }or, more technically,
the conditions fostering a satisfying life, can be attributed in part
to the existence of large international surveys which ask the same
life evaluation question, suitably translated into local languages,
across starkly different countries and cultures \parencite{Barrington-Leigh-WHR2022-trends-conception-wellbeing,Helliwell-Wang-WHR2012-chapter-2}.
The World Happiness Report has, annually since 2012, reported a ranking
across nearly 160 countries in the average answer to a single life
evaluation question from one such survey, the Gallup World Poll \parencite[e.g., ][]{Helliwell-et-al-WHR2025-chapter-2}.
This paper begins by addressing a straightforward question: are those
rankings reproducible, for instance by running a similar survey, or
by asking a slightly different question?

In principle, the rankings are not an object of primary interest to
researchers in the field. Due for instance to the desirable convergence
of transition or developing countries catching up to richer ones in
various ways, one may expect rankings of some more developed countries
to go down even when their average reported numerical life evaluations
are going up. Academic researchers are more interested in identifying
causal effects of policy environments and other circumstances, changes,
and choices on populations' life evaluations. These effects are inferred
by estimating coefficients in models that explain differences and
changes in life evaluations at the individual level or averaged over
groups. Such estimates are thought to be a key ingredient to begin
detailed policy design for wellbeing at different levels \parencite{Frijters-Clark-Krekel-Layard-BPP2020-Happy-Choice-SWB-as-goal-for-government,happiness-research-institute-2020-WALYs,Frijters-Krekel-2021-SWB-policy-handbook,UK-Treasury-2021-GreenBook-supplement-wellbeing,WhatWorksWellbeing-report2024-systematic-rapid-reviews-DOHC,Frayman-Krekel-Layard-MacLennan-Parkes-DRAFT2024-value-for-money}.

On the other hand, the researcher's ideal can be elusive. Not all
salient circumstances change on observable time scales or vary independently
of other circumstances. Therefore, population-wide values of a summative,
multi-faceted metric like a life evaluation cannot be fully accounted
for by specific, disaggregated causes and marginal effects. As a result,
international rankings remain important for addressing the broadest
policy questions about the success of different political and economic
systems, as judged by the experience of their populations. For instance,
knowledge of the high life evaluations in Nordic countries has brought
attention to their overall policy environment, lending credence to
the idea that a high trust, cohesive, supportive, high productivity,
individualist, social democratic society is the best model so far
for generating a happy populace.

At the same time, some have questioned whether the subjective life
evaluation approach could somehow be biased in favor of Western respondents
\parencite{Lomas-et-al-Chapter6WHR2022-balance-harmony}. This would
contradict the widespread assumption among economists in the field
that responses are internationally and interculturally comparable.
One problem could be cultural differences in conceptions of wellbeing
\parencite{Diener-Oishi-Lucas-ARP2003,Diener-Suh-2003-culture-and-SWB,Oishi-Graham-Kesebir-Galinha-PSPB2013-meaning-of-happiness-over-time,Oishi-chapter2010}.
Early work on life evaluations from the Gallup World Poll  suggested
strong comparability around the world, in the sense that similar coefficient
estimates for the correlates of life satisfaction were found in regions
and countries around the world \parencite{Helliwell-Barrington-Leigh-Harris-Huang-2010}
and that cultural effects play a limited role in explaining cross-country
differences compared with objective life circumstances \parencite{Exton-Smith-Vandendriessche-OECD2015}.
Findings in such studies imply that the life evaluation question taps
into something universal about human experience, a premise which
 underlies the country rankings of the prominent annual World Happiness
Reports. 

Another possible problem relates to the reporting function rather
than problems of translating the question. Using surveys with a battery
of subjective questions rather than one key one, Likert-scale responses
have been explained in part based on individual tendencies towards
\emph{moderate responses} (the central option) or \emph{extreme responses}
\parencite[top and bottom options; see][]{Hamamura-Heine-Paulhus-PID2007-cultural-differences-response-styles,Khorramdel-vonDavier-Pokropek-BJMSP2019-extreme-response-styles-model},
as well as to norms related to the appropriateness of particular feelings
\parencite{Diener-Oishi-Lucas-ARP2003}. In such models, psychologists
have emphasized the role of \emph{personality} and \emph{culture}
in shaping subjective wellbeing reports \parencite[e.g.,][]{Diener-Oishi-Lucas-ARP2003,Oishi-chapter2010,Heine-Lehman-Peng-Greenholtz-JPSP2002,Hamamura-Heine-Paulhus-PID2007-cultural-differences-response-styles}.
Although this branch of \emph{item response theory is }not directly
applicable to the interpretation of individual questions with largely-numeric
response scales as used in many life evaluation questions, a related
approach models the \emph{focal value rounding} tendency of individuals
answering just one life evaluation question, based on their known
personal characteristics, such as education \parencite{Barrington-Leigh-JPubEcon2024-focal-values}.
A number of other recent studies have begun to reexamine the possibility
of empirically significant bias due to differences in reporting functions
\parencite{Bond-Lang-JPE2019,Kaiser-Vendrik-DRAFT2023-2023much,Oparina-Srisuma-JBES2022-measurement-error-SWB-UK,Goff-DRAFT2025-identification-ordinal-outcomes}.
Such bias would be a threat to the possibility to rank subgroups'
mean life evaluations, as well as to make inference about the determinants
of quality of life.

Despite these efforts, no obvious problematic pattern in responses
has been isolated and found to be large enough to disrupt the global
rankings, especially at its top. Moreover, country averages of life
satisfaction tend to be relatively well explained by differences in
other measured circumstances which are \emph{not} thought to be subject
to the same potential problem \parencite{Exton-Smith-Vandendriessche-OECD2015}.

This paper presents evidence of striking gaps in our understanding
of international rankings of life evaluations. It begins by revisiting
a comparison of two forms of life evaluation questions in the Gallup
World Poll. The comparison is then made for a recent survey, the Global
Flourishing Study, which poses the same two questions. These four
measures are compared with one life evaluation question from a third
source, the World Values Survey. \secref{Data} describes these surveys,
the life evaluation questions, and the construction of matched samples.
\secref{Country-ranks} presents pairwise comparisons of country ranks
and argues that these qualitatively refute seven possible, relatively
straightforward hypotheses about why incongruities might arise when
comparing ranks from different questions or surveys. \secref{Univariate-response-distribution}
investigates the shape of response distribution in the 10- or 11-
point scales, finding further evidence of nontrivial problems in the
reproducibility of distributions of national samples of life evaluations.
\secref{Joint-response-distributions} shows examples of the bivariate
(joint) response distributions from individuals who answered two life
evaluation questions in the same survey, and \secref{Model-inference}
returns to the approach of modeling responses in order to compare
estimated coefficients on predictors across two life evaluation questions.
\secref{Discussion-and-Conclusion} concludes.

\section{Data\protect\label{sec:Data}}

Three survey sources provide the data presented below: Wave 4 (1999\textendash 2004),
Wave 5 (2005\textendash 2009), Wave 6 (2010\textendash 2014), and
Wave 7 (2017\textendash 2022) of the \textbf{World Values Survey}
\parencite[{\bfseries WVS}, 76 countries; see][]{world-values-survey-1981-2020}
conducted in one calendar year per wave in each country; annual waves
of the repeated cross-section \textbf{Gallup World Poll} \parencite[{\bfseries GWP}, 157 countries; see][]{Gallup-methodology-2012,GALLUP2014-Cantril-Scale}
from 2006-{}-2022; and the first wave of the \textbf{Global Flourishing
Study} \parencite[{\bfseries GFS}, 22 countries; see][]{GFS-Global-Flourishing-Study-Data-2024}
in 2023.

The WVS poses the \textbf{life satisfaction question} (\textbf{LS})
on a 1\textendash 10 scale, while the other two surveys use 0\textendash 10
scales for both LS and \textbf{Cantril ladder} (\textbf{CL}). The
GFS poses both questions to all respondents. However, the GWP, which
always includes CL, only fielded the LS question between 2007 and
2010, and only in a subset of countries. Indeed, only 114 countries
ever received both questions in the GWP, and, except for five (Belgium,
Belarus, Denmark, Sri Lanka, and Singapore) which received LS twice,
each received it only once. The number of countries with LS as well
as CL in the GWP were 37 in 2007, 67 in 2008, and just 9 in 2009 and
6 in 2010.

\subsubsection*{Matched pairs}

In order to remove time variation as much as possible in what follows,
matched pairs are constructed for comparisons of life evaluation samples
across countries.

For comparing LS and CL from the GWP, all individuals who have answered
both are selected for each country. As mentioned above, these are
all in one year per country except for five countries. When WVS (LS)
is compared with CL from GWP, only exact year matches are used. When
LS is compared between WVS and the GWP, the following approach is
used: for each country available in the WVS, the closest WVS year
available to the GWP year with LS responses is chosen. When WVS (LS)
is compared with either measure from the GFS, the most recently available
wave from WVS is used for each country in common. For comparing CL
between the GWP and the GFS, the most recent year available from 2021
(5 countries) or 2022 (17 countries) is used for the GWP.

\subsubsection*{Question order}

Because of their superlatively open scope, and in order to minimize
framing effects, life evaluation questions are often placed at the
very beginning of a survey. In the GFS, the Cantril ladder question
opens the survey, phrased as follows:
\begin{quote}
Please imagine a ladder with steps numbered from zero at the bottom
to ten at the top. The top of the ladder represents the best possible
life for you, and the bottom of the ladder represents the worst possible
life for you. On which step of the ladder would you say you personally
feel you stand at this time? {[}0=Worst possible, 10=Best Possible{]}
\end{quote}
Immediately following it is a question on the same scale about life
in 5 years, followed by a question about happiness, followed by the
life satisfaction question:
\begin{quote}
Overall, how satisfied are you with life as a whole these days? {[}0=Not
satisfied with your life at all, 10=Completely satisfied with your
life{]} 
\end{quote}
However, according to the questionnaire, the interviewer introduces
the happiness and life satisfaction questions, along with several
subsequent ones, by reiterating the vertical ladder image for use
in the response scale: ``Now continue to think about the ladder with
the top of the ladder at ten being the best possible state or arrangement
and the bottom of the ladder at zero being the worst possible state
or arrangement'' \parencite{GFS-Global-Flourishing-Study-Data-2024}.
This is an unusual approach.

In the GWP, the survey opens with four questions about household geography
and size, followed by a dichotomous satisfaction question, and then
the Cantril ladder, worded nearly identically to the GFS version.
By contrast with the GFS, when the life satisfaction question was
fielded, it appeared near the end of the survey, just before questions
on household income. It is phrased:
\begin{quote}
All things considered, how satisfied are you with your life as a whole
these days? Use a 0 to 10 scale, where 0 is dissatisfied and 10 is
satisfied.
\end{quote}
In the WVS, life satisfaction is asked near the end of the first long
module, roughly 20\% of the way through the interview, and phrased
very similarly to that of the GWP:
\begin{quote}
All things considered, how satisfied are you with your life as a whole
these days? Please use this card to help with your answer. {[}1 `Dissatisfied',
\dots   10 `Satisfied'{]}
\end{quote}

\subsubsection*{Other variables}

In \secref{Model-inference}, survey questions are included in predictive
models of life evaluations. Each asks about an objective fact concerning
respondents' lives. These variables are detailed in Appendix \tabref{GFS-descriptive-stats}.

\section{Country ranks\protect\label{sec:Country-ranks}}

\begin{table}
  \begin{tabular}{{lcllp{.15\textwidth}rrr}}
\hline\hline
 Surveys    & Metrics   & $r$(ranks) & $r$(indiv) & Fraction {$\Delta$$>$1 quartile} %
&   N &\\ \hline
 WVS (\yearRangeLSWVSmatchedtoGWP) vs  GWP (\yearRangeCLGWPmatchedtoWVS)& LS vs CL &       0.66 &  &                       0.37 &  76        \\
 GWP (\yearRangeLSGWP)                                                    & LS vs CL &       0.92 & \GWPindivCorrLSCL &                        0.06 & 124 &       \\ %
 GWP (GFS countries, \yearRangeCLGWP)                                      & LS vs CL &       0.94 & \GWPindivCorrLSCLGFScountries &                        0 & 22 &       \\
 GFS (\yearRangeCLGFS)                                                      & LS vs CL &       0.59 & \GFSindivCorrLSCL  &                        0.27 &  22 &       \\
 GWP (\yearRangeCLGWPTwentyOneTwentyTwoGFScountries) vs GFS (\yearRangeCLGFS)  & CL        &       0.80 & &                        0.14 &  21 &       \\
 GWP (\yearRangeLSGWPGFScountries) vs GFS (\yearRangeLSGFS)& LS       &       0.23 & &                        0.59 &  17 &       \\
 WVS (\yearRangeLSWVSmatchedtoGFS) vs GFS (\yearRangeCLGFS)& LS vs CL        &       0.75 & &                        0.19 &  21 &       \\
 WVS (\yearRangeLSWVSmatchedtoGFS) vs GFS (\yearRangeLSGFS)& LS      &       0.56 &  &                       0.24 &  21 &       \\
 WVS (\yearRangeLSWVSmatchedtoGWP) vs GWP (\yearRangeLSGWP)& LS      &       0.69 &  &                       0.28 &  74 &  \\ \hline \hline
  \end{tabular}

\caption[Summary of correlations for rank comparisons]{Summary of correlations for rank comparisons. The table lists survey
year ranges resulting from each match, the correlation of ranks, the
fraction of countries for which ranks differ by more than a quartile,
and number of countries in the matching pair, and, for the surveys
which posed both questions, the correlation across individuals between
LS and CL reported values.\protect\label{tab:Summary-of-rank-comparisons}}
\end{table}

There are several sensible pairwise comparisons of country rankings
among the five life evaluation sources described above. These comparisons
are summarized in \tabref{Summary-of-rank-comparisons} and detailed
below.

\subsection*{The World Values Survey and Gallup World Poll}

To start, \figref{LSWVS-vs-CLGWP-ranks} compares the rank order of
country means from the two most prominent global surveys of life evaluations.
This is a comparison of different questions from different surveys:
the WVS' life satisfaction (LS) question and the GWP's Cantril ladder
(CL). The overall correlation between the ranks is 0.66, yet there
are very stark inconsistencies. The group of countries that has been
somewhat famously at the top of the World Happiness Report's annual
ranking is not at the top according to the life satisfaction question
in WVS. Moreover, there appear to be systematic differences across
cultural groups between the two data sets. Colors show \citeauthor{Inglehart-Baker-ASR2000}'s
\parencite*{Inglehart-Baker-ASR2000} classification of countries
into nine cultural groups. \tabref{LSWVS-vs-CLGWP-differences-by-cultgroup}
quantifies these systematic differences across country groups.\footnote{The means and confidence intervals for differences in ranks are calculated
by Monte Carlo  simulation, assuming normally distributed beliefs
about the mean of each country.} The difference in means is small for English Speaking countries and
Protestant Europe, slightly higher for Catholic Europe, increasingly
higher for Orthodox, Confucian, and Islamic, and truly large ($\geq1.7)$
for Latin America, Africa, and South Asia. 

\afterpage{\newpage
\centering
\begin{figure}
\includegraphics[width=1\columnwidth]{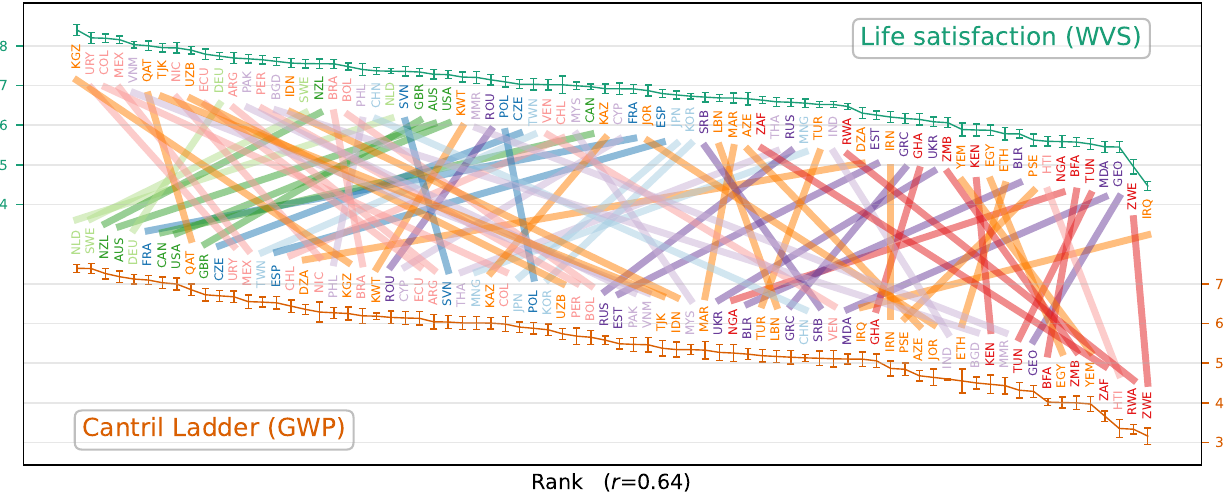}

\caption[Country ranks: life satisfaction (WVS) and Cantril ladder (GWP)]{Comparison of country ranks using mean life satisfaction from the
World Values Survey (WVS) and mean Cantril ladder from the Gallup
World Poll (GWP). Countries are labelled using their ISO3 codes, and
colored according to the cultural groupings shown in \tabref{LSWVS-vs-CLGWP-differences-by-cultgroup}.
The range of source years is shown in \tabref{Summary-of-rank-comparisons}.\protect\label{fig:LSWVS-vs-CLGWP-ranks}}
\end{figure}

\begin{table}
\begin{centering}
Life satisfaction (WVS) versus Cantril ladder (GWP)
\par\end{centering}
\begin{centering}
\centering 
\includegraphics{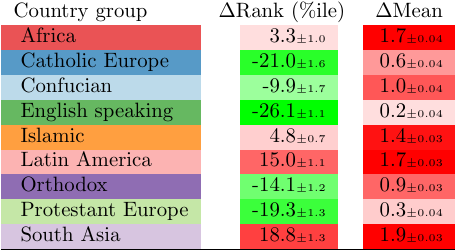}

\par\end{centering}
\caption[Average differences by cultural group for \figref{LSWVS-vs-CLGWP-ranks}]{ Average country rank differences by cultural group for life satisfaction
responses from WVS, as compared with Cantril ladder from the GWP.
Positive $\Delta$Rank means a higher ranking. See \figref{LSWVS-vs-CLGWP-ranks}
for rankings. \protect\label{tab:LSWVS-vs-CLGWP-differences-by-cultgroup}}
\end{table}

\clearpage\newpage

\begin{figure}
\includegraphics[width=1\textwidth]{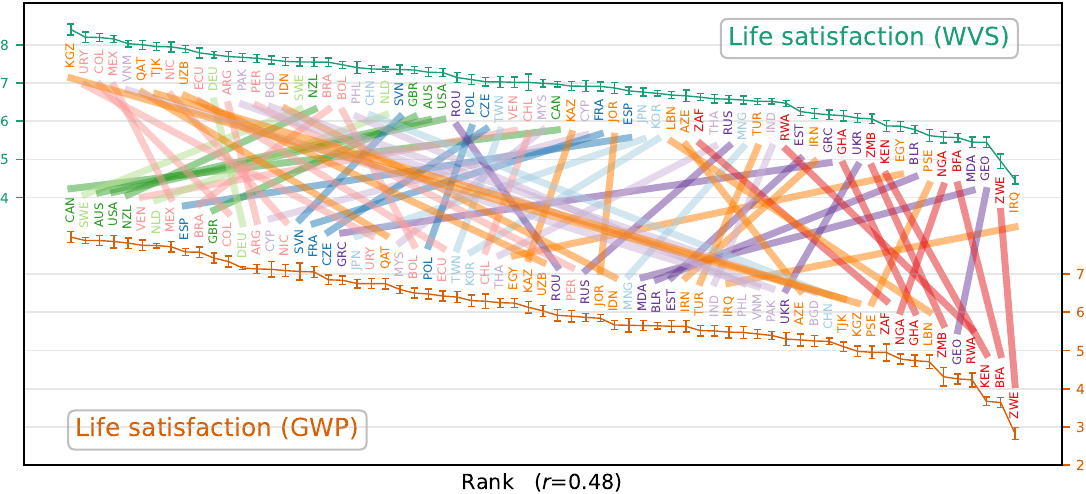}

\caption[Country ranks: life satisfaction from the WVS and GWP]{Comparison of country ranks using mean life satisfaction from the
WVS versus the GWP. \protect\label{fig:LSWVS-vs-LSGWP-rankings}}
\end{figure}

\begin{table}
\begin{centering}
Life satisfaction: WVS versus GWP
\par\end{centering}
\begin{centering}

\includegraphics{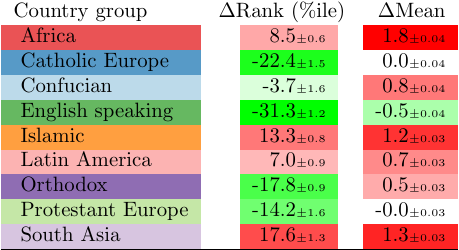}
\par\end{centering}
\caption{Average differences by cultural group for \figref{LSWVS-vs-LSGWP-rankings}\protect\label{tab:LSWVS-vs-LSGWP-differences-by-cultgroup}}
\end{table}

\clearpage\newpage

\begin{figure}
\includegraphics[width=1\textwidth]{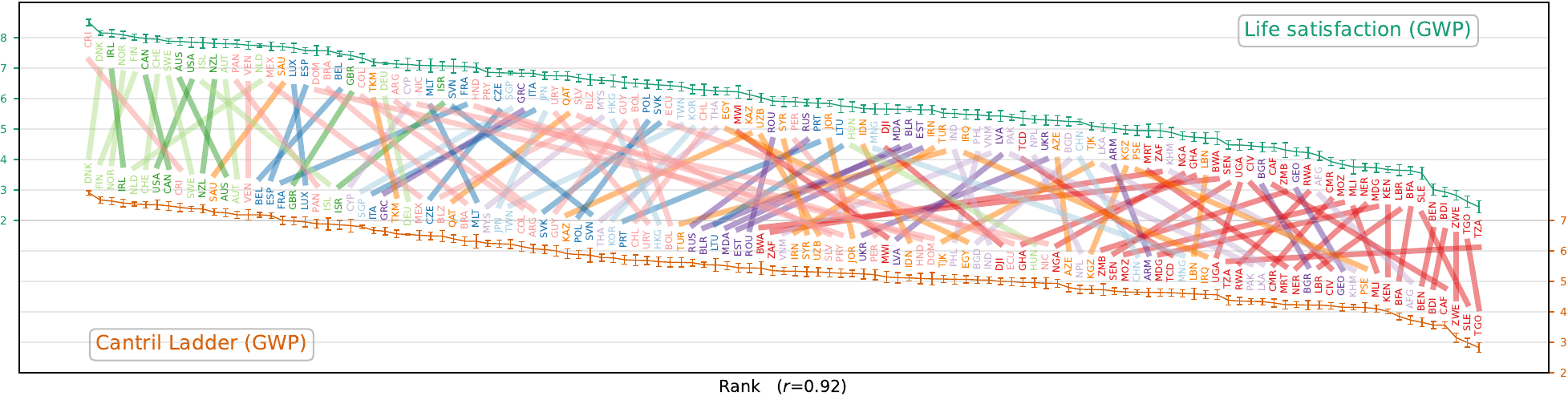}

\caption[Country ranks: GWP life satisfaction and Cantril ladder]{Comparison of country ranks using mean life satisfaction and mean
Cantril ladder, both from the GWP.\protect\label{fig:LSGWP-vs-CLGWP-rankings}}
\end{figure}
\begin{table}
\begin{centering}
Life satisfaction versus Cantril ladder, both from GWP
\par\end{centering}
\begin{centering}

\includegraphics{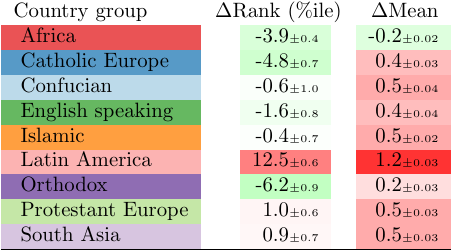}
\par\end{centering}
\caption{Average differences by cultural group for \figref{LSGWP-vs-CLGWP-rankings}\protect\label{tab:table-cultgroup-mean-rankdiffs-LSGWP-vs-CLGWP-standalone}}
\end{table}

}

Before investigating what might be leading to some of the extreme
discrepancies, by looking at response distributions within a country,
let us see how other surveys and questions compare. The disagreements
in rank shown in \figref{LSWVS-vs-CLGWP-ranks} could in principle
be due to differences in how the \emph{surveys} are implemented (question
order, framing, interview mode, sampling, etc) or in how people interpret
and respond to the two different \emph{questions}. 

To address the former possibility, \figref{LSWVS-vs-LSGWP-rankings}
compares rankings calculated from responses to the \emph{same} LS
question but asked in the two different repeated survey series of
\figref{LSWVS-vs-CLGWP-ranks}. The matching of survey years in \Figref{LSWVS-vs-LSGWP-rankings}
is not as precise as in \figref{LSWVS-vs-CLGWP-ranks}; see \nameref{sec:Data}.
Overall, the correlation of ranks across countries, 0.69, is very
similar and moderately strong, yet the disagreements in rank order
are once again dramatic and statistically significant and again vary
by cultural group. \tabref{LSWVS-vs-LSGWP-differences-by-cultgroup}
shows these systematic differences across groups.  In addition to
the systematic differences in rank, African countries responded in
WVS on average 1.8 higher on the 0\textendash 10 scale than when faced
with the same question in GWP.

Could both \figref{LSWVS-vs-CLGWP-ranks} and \ref{fig:LSWVS-vs-LSGWP-rankings}
be explained instead by differences in the administration of the \emph{surveys}?
Comparing responses to the two questions answered by the same individuals
in the same survey is possible for the GWP using the years 2007\textendash 2010,
and the resulting country rankings are compared in \figref{LSGWP-vs-CLGWP-rankings}.
Here the correlation of country ranks, 0.92, is much higher, indicating
relative consistency, yet some individual countries, and the majority
of the Latin American ones, still exhibit large systematic differences
(\tabref{table-cultgroup-mean-rankdiffs-LSGWP-vs-CLGWP-standalone}).
In light of the enormous public attention given to the country rankings
in the WHR, these discrepancies are not likely to sit well with countries
placing much lower than they would under an alternative measure.

\subsection*{The Global Flourishing Study}

\begin{figure}
\includegraphics{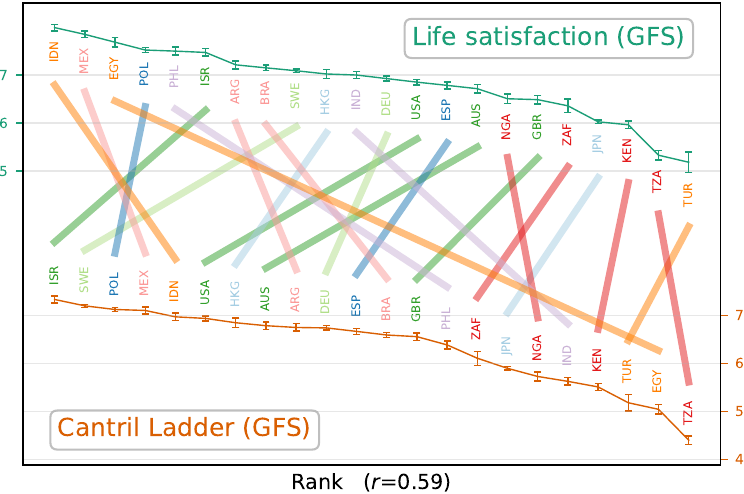}

\caption[Country ranks: GFS life satisfaction and Cantril ladder]{Life satisfaction versus Cantril ladder, both from GFS.\protect\label{fig:LSGFS-vs-CLGFS-rankings}}
\end{figure}

In 2024 data from the first wave of a new global survey, the Global
Flourishing Study (GFS), became available. The GFS also fielded both
LS and CL life evaluation questions. \figref{LSGFS-vs-CLGFS-rankings}
compares ranks for the 22 countries in the GFS, when average life
evaluations are calculated using the two different questions. In this
case, the correlation of ranks at the country level is only 0.57.
With fewer countries, generalizing across cultural or other groups
of countries is harder, but in \tabref{table-cultgroup-mean-rankdiffs-LSGFS-vs-CLGFS}
we see that there are nevertheless statistically significant differences
by country group. Undoubtedly, there are some stark differences, including
that of Egypt, which is near the top of the LS scale but near the
bottom of the CL scale.

\begin{figure}
\includegraphics{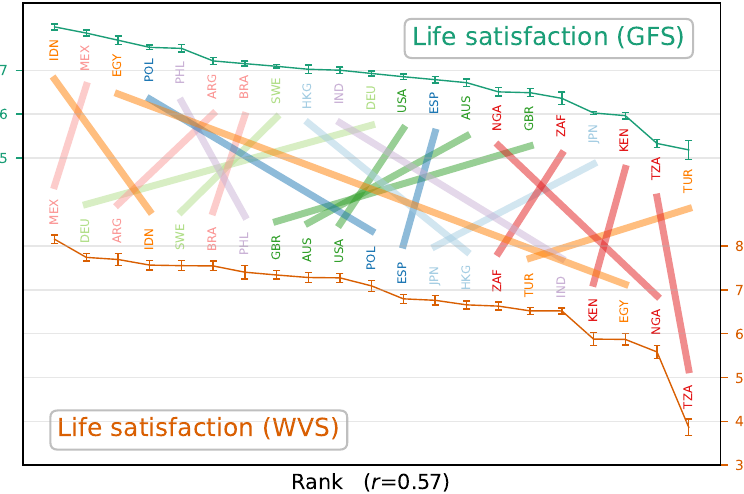}

\caption[Country ranks: life satisfaction from WVS and GFS]{Rankings calculated using life satisfaction from WVS versus GFS.
See \tabref{table-cultgroup-mean-rankdiffs-LSWVS-vs-LSGFS} for average
differences by country group. \protect\label{fig:LSWVS-vs-LSGFS-rankings}}
\end{figure}

In the absence of other evidence, \figref{LSGFS-vs-CLGFS-rankings}
might once again lead one to form the hypothesis that differences
in the interpretation of the two life evaluation questions are behind
the incongruity of ranks. Under this explanation, differences in ranks
would reflect substantial differences across countries in certain
dimensions of life which also mattered differently for determining
answers to the two questions. However, a straightforward interpretation
of this sort can again be rejected using GFS data, as it was above
for the WVS \textendash{} GWP pair. Using the most recent wave of
the WVS to compare with GFS's 2023 life satisfaction responses (\figref{LSWVS-vs-LSGFS-rankings}),
we find a nearly identical, low correlation of ranks and equally stark
discrepancies at the individual country level, even though in this
case both rankings are based on the life satisfaction question.

\begin{figure}
\includegraphics{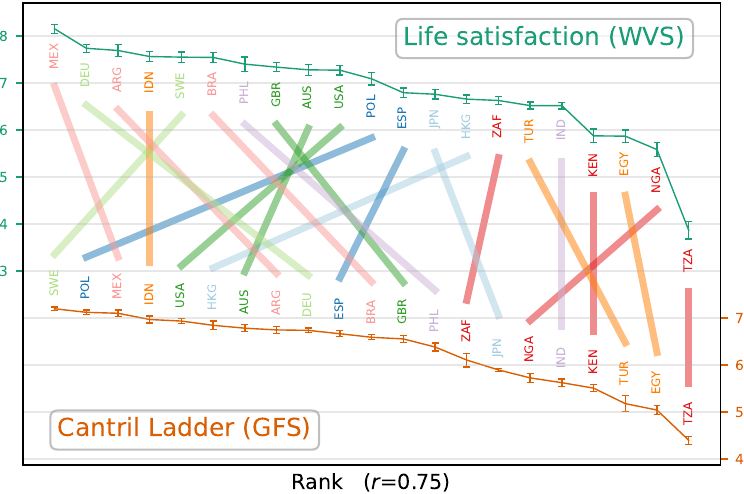}

\caption[Country ranks: life satisfaction (WVS) and Cantril ladder (GFS)]{Ranks calculated from life satisfaction (WVS) versus Cantril ladder
(GFS). See \tabref{table-cultgroup-mean-rankdiffs-LSWVS-vs-CLGFS}
for average differences by country group.\protect\label{fig:LSWVS-vs-CLGFS-rankings}}
\end{figure}

\begin{table}
\begin{centering}
Life satisfaction versus Cantril ladder, both from GFS
\par\end{centering}
\begin{centering}

\includegraphics{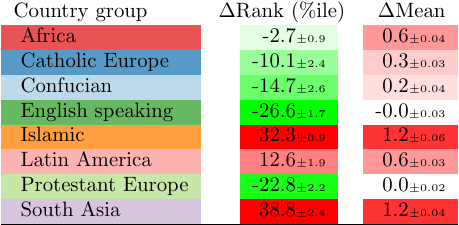}
\par\end{centering}
\caption{Average differences by cultural group for \figref{LSGFS-vs-CLGFS-rankings}\protect\label{tab:table-cultgroup-mean-rankdiffs-LSGFS-vs-CLGFS}}
\end{table}
Even more remarkably, when life satisfaction from WVS is compared
with Cantril ladder (rather than life satisfaction) from GFS, the
agreement is better, with a correlation of 0.75 (\figref{LSWVS-vs-CLGFS-rankings}).
Egypt, which is one of the extreme cases in Figures \ref{fig:LSGFS-vs-CLGFS-rankings}
and \ref{fig:LSWVS-vs-LSGFS-rankings}, is, in \figref{LSWVS-vs-CLGFS-rankings},
in good agreement across these differing measures on different surveys.

\begin{figure}
\includegraphics{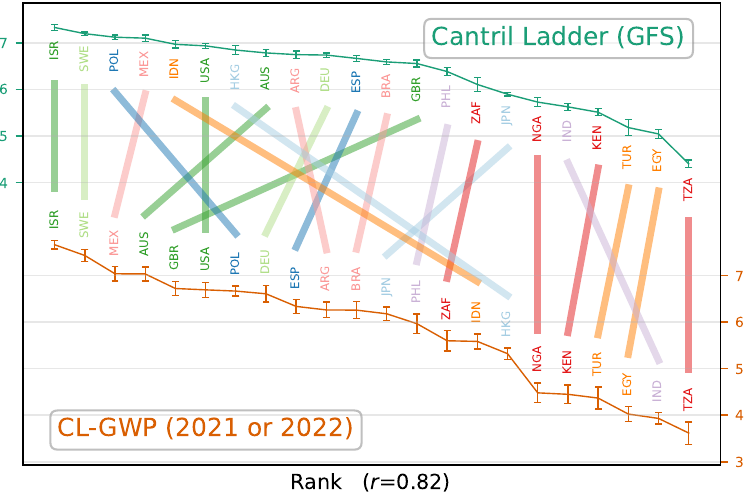}

\caption[Country ranks: Cantril ladder from GFS and GWP]{Ranks calculated from Cantril ladder (GFS versus GWP).\protect\label{fig:CLGWP-vs-CLGFS-rankings}}
\end{figure}

One final comparison is shown in \figref{CLGWP-vs-CLGFS-rankings}.
When the Cantril ladder responses are compared from GWP and GFS, the
agreement is relatively good, with an overall correlation of 0.80.
This is the kind of consistency one might hope for when publishing
international rankings of these measures, although even if this were
the full picture one would need to investigate the several statistically
significant exceptions.

\begin{table}
\begin{centering}
Cantril ladder (GWP versus GFS)
\par\end{centering}
\begin{centering}

\includegraphics{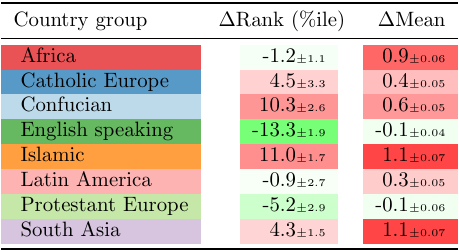}
\par\end{centering}
\caption{Average differences by cultural group for \figref{CLGWP-vs-CLGFS-rankings}\protect\label{tab:table-cultgroup-mean-rankdiffs-CLGWP-vs-CLGFS}}
\end{table}

\subsection*{Discussion}

We might characterize the preceding evidence as addressing the following
hypotheses:
\begin{enumerate}
\item Major life evaluation metrics are generally consistent with each other
at the country level.
\item The two major life evaluation questions tap into different characteristics
of life to different degrees, but consistently across surveys.
\item The two major life evaluation questions are interpreted differently
across countries, but consistently across surveys.
\item The three major surveys have differences in implementation and survey
content or framing which induce distinct but consistent impacts on
any life evaluation questions.
\item The two major life evaluation questions tap into different characteristics
of life to different degrees, but when both life evaluation questions
are posed in the same survey, that which is posed first will have
a dominant influence on the responses to both.
\item Large differences in life evaluations reported across surveys can
usually be explained by the countries in question undergoing rapid
change, coupled with subtle differences in survey timing.
\end{enumerate}
The first is rejected by any of Figures \ref{fig:LSWVS-vs-CLGWP-ranks}\textendash \ref{fig:CLGWP-vs-CLGFS-rankings}.
Hypothesis 2 is rejected by the strong disagreements shown in \figref{LSWVS-vs-LSGWP-rankings}
and \figref{LSWVS-vs-LSGFS-rankings} and, arguably, the weaker disagreements
in \figref{CLGWP-vs-CLGFS-rankings}. Hypothesis 3 is rejected by
the same evidence, as cultural or country-specific effects qualify
as unmeasured life conditions. If there was a problem translating
the questions, or different conceptions of the two questions across
countries, then we might still expect to see consistency across surveys
asking the same question with the same population sampling frame.
On the contrary, we see stark differences across surveys.

Hypothesis 4 predicts that GWP and GFS, which have fielded both questions
in the same survey, should find consistent results from the two questions.
This is patently false in the GFS (\figref{LSGFS-vs-CLGFS-rankings})
and, to a somewhat lesser degree, the GWP (\figref{LSGWP-vs-CLGWP-rankings}),
which exhibits a strong overall correlation despite some coherent
cultural differences.

One way in which those surveys could differ is in the order of and
framing around the two questions. However, if one question influences
the other, it does not seem to be in any simple way as suggested in
Hypothesis 5. In both the GFS and the GWP, CL is asked before LS.
The GFS poses the LS question nearly immediately after the CL question,
and indeed even refers to using the same scale to answer about LS.
In spite of this, the answers in the GFS disagree more strongly than
in the GWP, which posed the CL question near the beginning and the
LS question near the end of the survey, with only questions about
income and some demographics following it. Below I investigate the
joint distribution of responses across individuals more explicitly.

\begin{figure}
\begin{centering}
\includegraphics[width=1\textwidth]{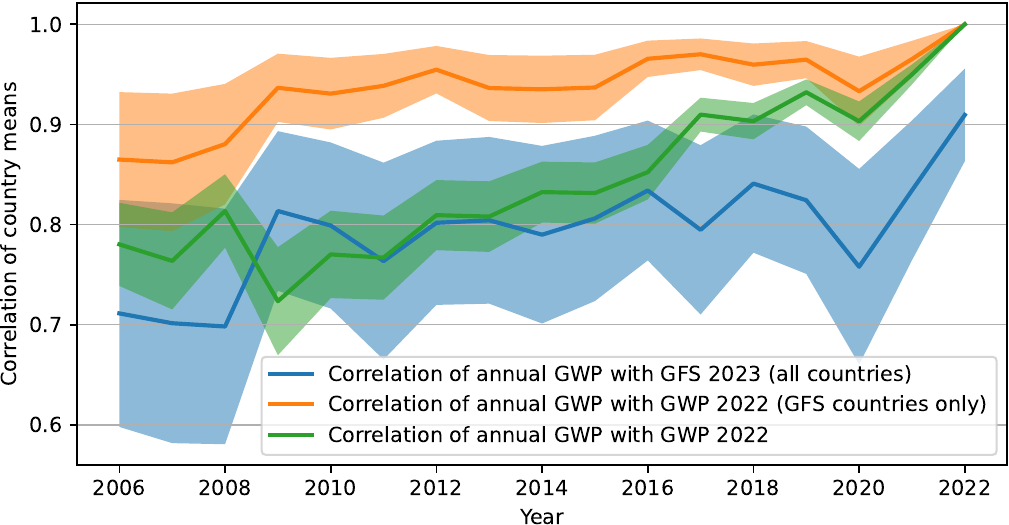}
\par\end{centering}
\caption[Consistency of country means, and of survey differences, across time
(GWP and GFS)]{Consistency of country means, and of survey differences, across time.
The green trace shows the correlation of each year's Cantril ladder
country means with those from 2022. The orange trace is the same but
restricted to the 20 countries surveyed in the GFS. The blue trace
is the correlation of those GWP annual country means with the GFS
mean responses of 2023. The high values of the orange trace prior
to 2022 show that measured means are highly reproducible. Comparing
the blue and orange traces shows that GWP values are consistently
different from those of GFS.\protect\label{fig:GWP-GFS-autocorrelation}}
\end{figure}

Hypothesis 6 represents an objection that even if two surveys are
carried out within the same calendar year, outlier results could arise
due to rapid changes in circumstances within countries. To better
address the temporality of the discrepancies highlighted so far, \figref{GWP-GFS-autocorrelation}
shows that differences between survey measures are persistent over
time. Focusing on the 22 GFS countries, the orange and blue traces
show that past GWP country means have been consistently different
from the recent GFS values, but consistently similar to the recent
GWP values. Thus, the scale of discrepancies across surveys is larger,
and broader, than the changes in mean country responses over time.

In light of the rejection of (1) \textendash{} (6), it appears that
any explanation of the non-reproducibility of country ranks must involve
the interpretation or reporting behavior varying not only between
the two wordings of questions, but \emph{simultaneously} across survey
contexts and \emph{simultaneously} across countries.

\section{Univariate response distributions\protect\label{sec:Univariate-response-distribution}}

The straightforward evidence above is based on mean responses at the
country level, and appears to reject any simple explanation of the
differences between life satisfaction and the Cantril ladder. A next
natural step in descriptive evidence is to look at the individual
response distribution and how it varies across questions, surveys,
and countries.

\afterpage{\clearpage\begin{landscape}

\begin{figure}
\centering{}\vspace{-10mm}\includegraphics[width=1\paperwidth]{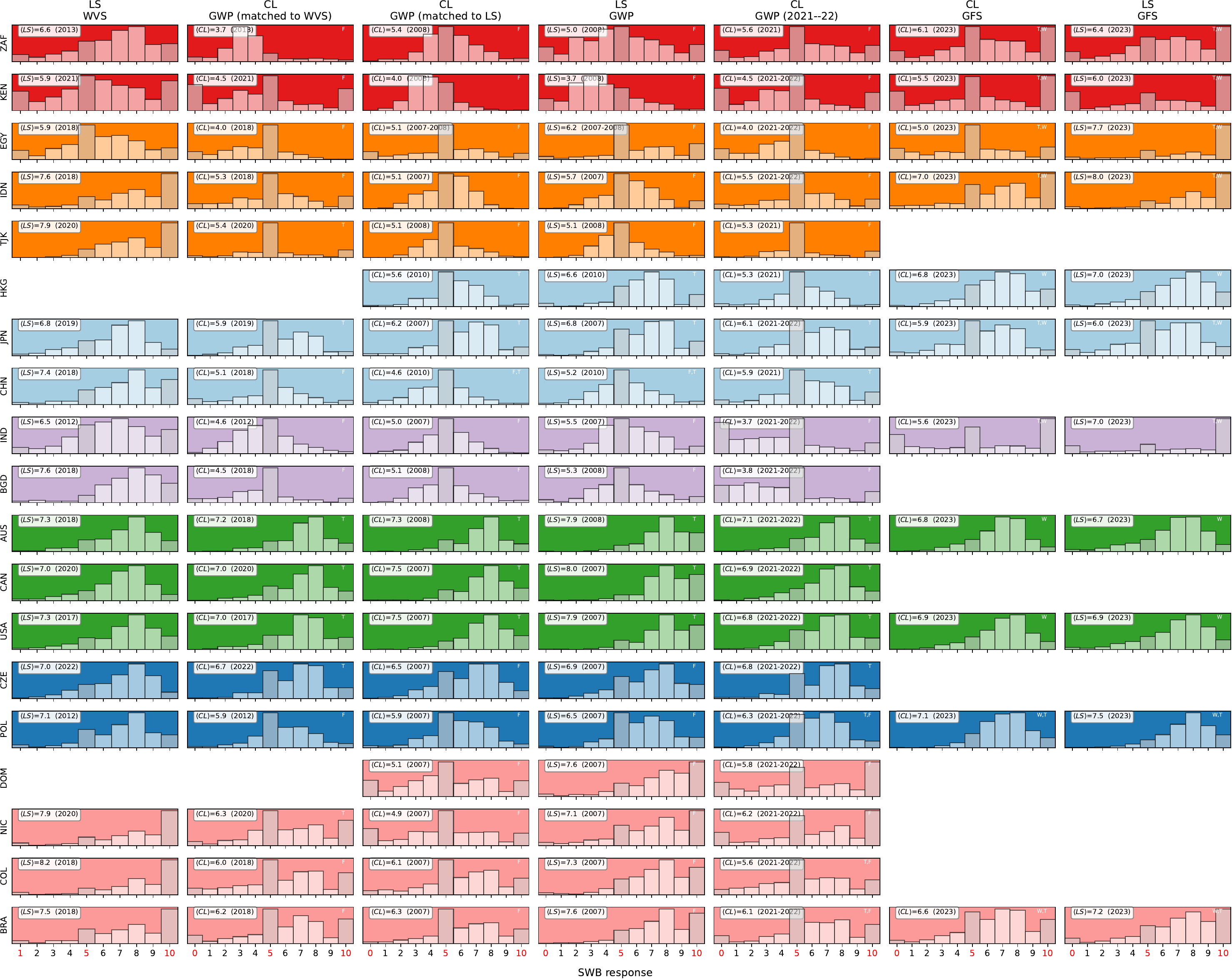}\caption[Response distributions for a selection of countries, for each survey
and question.]{Response distributions to life satisfaction (LS) and Cantril ladder
(CL)  from the WVS, GWP, and GFS, for a selection of countries.\protect\label{fig:distributions-grid-by-country}}
\end{figure}

\clearpage\end{landscape}%
}%

Consider the first  pair of columns in \figref{distributions-grid-by-country},
which shows responses from a selection of countries, each notable
for at least one large discrepancy in the previous section. The distributions
are qualitatively different, with some systematic patterns across
countries. In most cases, the distribution for CL in the GWP is shifted
to the left as compared with that of LS in the WVS. In addition, in
most countries (but not in South Africa, and hardly in the English
speaking countries) there is a more pronounced enhancement of the
``5'' response for CL (GWP) than for LS (WVS). In some cases, this
is quite extreme, for instance in the Tajikistan case. These qualitative
differences lead to lower means for the CL, but the differences in
means (tabulated in \tabref{LSWVS-vs-CLGWP-differences-by-cultgroup})
vary considerably. For instance, it is an enormous +3.1 for Bangladesh,
+2.5 for Tajikistan, +2.4 for China, +2.3 for Indonesia, and +2.2
for Colombia, but only 0\textendash 0.1 for Canada and the USA. 

The tendency to prefer the ``5'' response seems in most cases to
coincide with a relative preference for the bottom (0 or 1) and top
(10) values as well. In fact one might hypothesize that the prevalence
of 5 in the CL (GWP) could be entirely explained by an overall lower
distribution of latent (i.e., true) wellbeing, coupled with the focal
value rounding (FVR) response function behavior identified and characterized
by \textcite{Barrington-Leigh-JPubEcon2024-focal-values}. For instance,
in Indonesia, both the CL (GWP) and LS (WVS) distributions show enhancements
both for 5 and for 10. However, with distributions centered around
5 and 7\textendash 8, respectively, there are few 10s for CL and few
5s for LS.

In some cases, however, there appear to be countries without much
FVR (South Africa) or where the enhancement for ``10'' seems less
strong in LS (WVS) than might be expected based on that of ``5''
in CL (GWP) \textemdash{} for instance, in Japan, Bangladesh, Czech
Republic, and Poland.

\figref{distributions-grid-by-cultgroup} shows the same columns but
with all countries now pooled across cultural groups. The patterns
in the CL (GWP) and LS (WVS) responses, described above for individual
countries, generalize to these pooled distributions.

Returning to \figref{distributions-grid-by-country}, the next two
columns compare LS and CL from matched years in the GWP. They show
that the qualitative differences just described are not intrinsic
to the question wording. That is, (as foreshadowed by \figref{LSGWP-vs-CLGWP-rankings})
response patterns are quite similar in the two questions when they
are asked together in the same survey. Although FVR varies markedly
across countries, it is exhibited similarly across this pair of questions.
However, there are exceptions even to this generalization \textemdash{}
for instance, the Latin American countries and maybe South Africa.
For Latin America, the prevalence of 5s in the CL (GWP) responses
and to a lesser degree, 10s in the LS (GWP) responses may account
for the larger average difference in means (\tabref{table-cultgroup-mean-rankdiffs-LSGWP-vs-CLGWP-standalone}
and \figref{distributions-grid-by-cultgroup}).

Some countries (Kenya, Indonesia, Tajikistan) exhibit qualitative
differences between the year matched to WVS (column 2) and that matched
to LS (column 3), though these are unusual. \figref{distributions-grid-over-time}
addresses the question of reproducibility of qualitative and quantitative
patterns across years within a single survey, and shows generally
stable or slowly changing distributions for CL in the GWP over time.

Turning next to the analogous comparison for GFS, columns 6 and 7
(the rightmost two columns) in \figref{distributions-grid-by-country}
compare responses to LS and CL, once again from the same individuals
in the same survey. The relationship between these two columns is
different from that of the corresponding pair for the GWP. There is
more of a shift to the right for LS in India and Egypt, as compared
with the wider distribution for CL. As a result of the very strong
FVR, this shift gives Egypt a stark difference (2.7) in means. The
strong FVR also leads to exceptionally high mean response for LS in
Egypt and Indonesia.  It is not clear by inspection whether the degree
of FVR can be said to vary across questions, for instance in India
and to some extent in South Africa and Brazil. \figref{distributions-grid-by-cultgroup}
shows that these shifts and strong FVR are features that generalize
to the rest of the cultural group for South Asia and Islamic countries.

By examining columns 5 and 6 of \figref{distributions-grid-by-country},
a final comparison may be made between CL in the GFS (in 2023) and
in the GWP (2021\textendash 2022). Some cases, like Japan, look remarkably
similar, while others, like Hong Kong and Indonesia, look qualitatively
dissimilar. In most countries except for the English Speaking group,
however, the use of ``10'' in CL responses is lower in the GWP than
in the GFS. These patterns are evident also in the pooled data of
\figref{distributions-grid-by-cultgroup}.

\section{Joint response distributions\protect\label{sec:Joint-response-distributions}}

\begin{figure}

\centering %
\begin{tabular}{m{1cm}m{7cm}m{7cm}m{1cm}}
\centering\rotatebox{90}{Argentina} & \includegraphics[width=68mm]{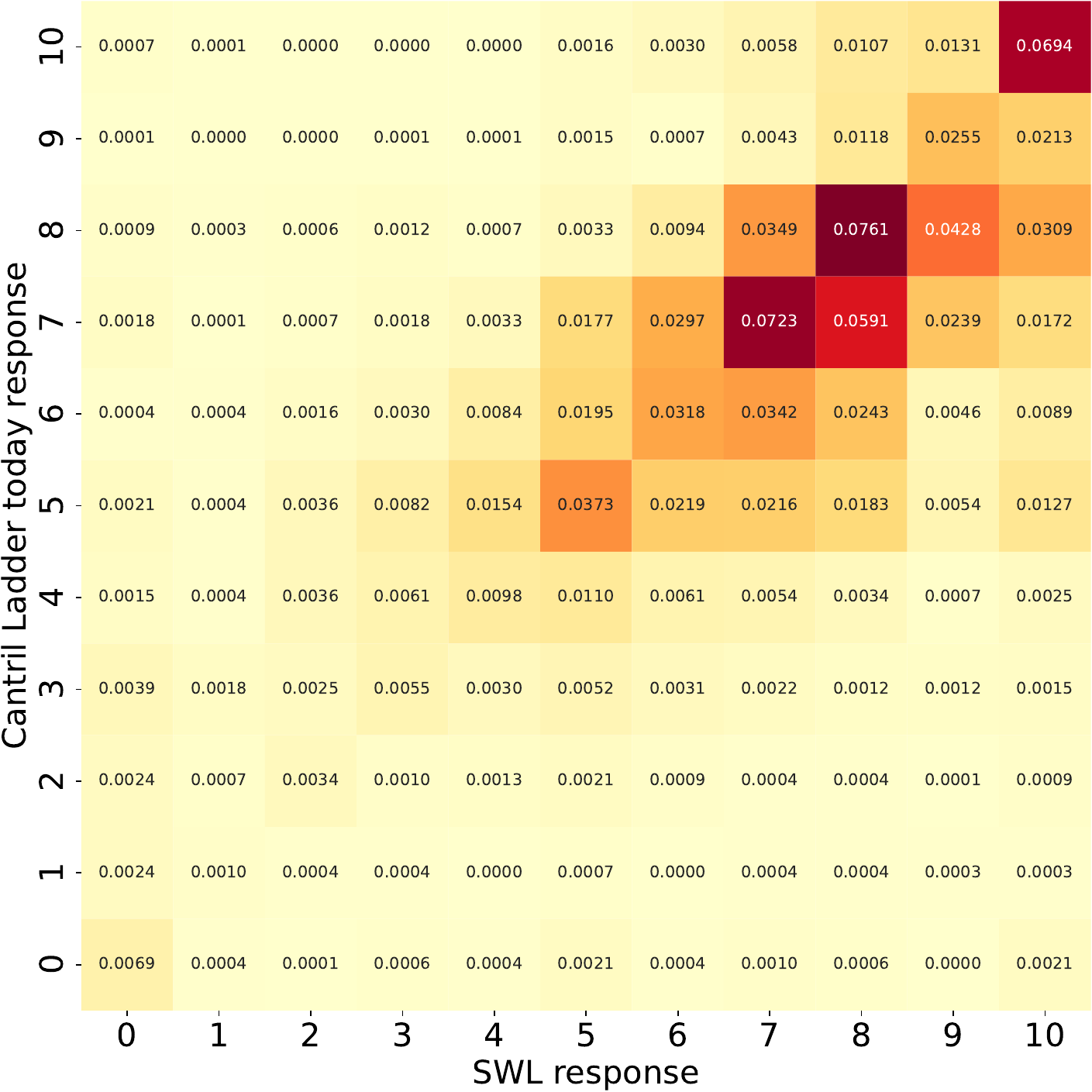} & \includegraphics[width=68mm]{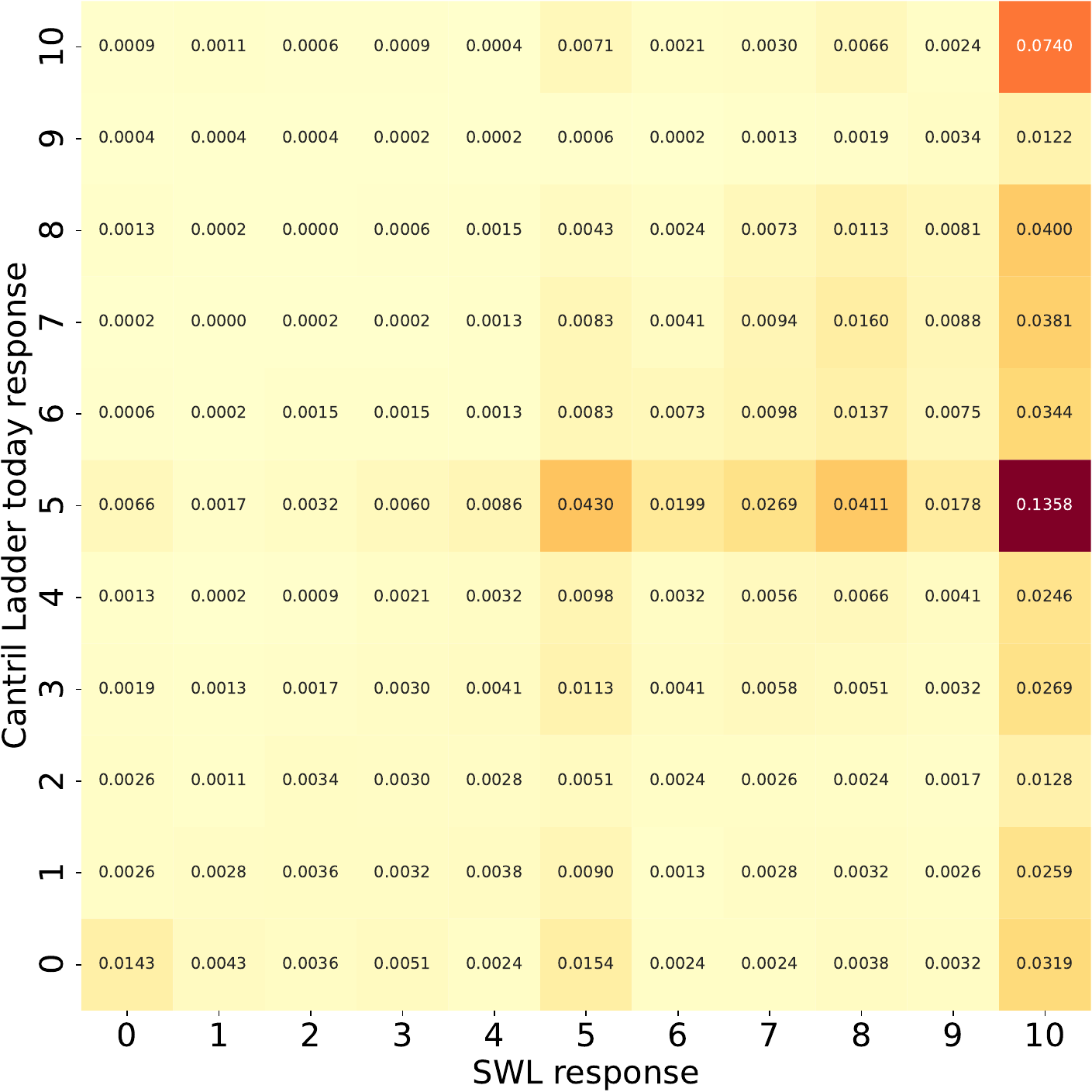} & \rotatebox{90}{Egypt}\tabularnewline
\centering\rotatebox{90}{Japan} & \includegraphics[width=68mm]{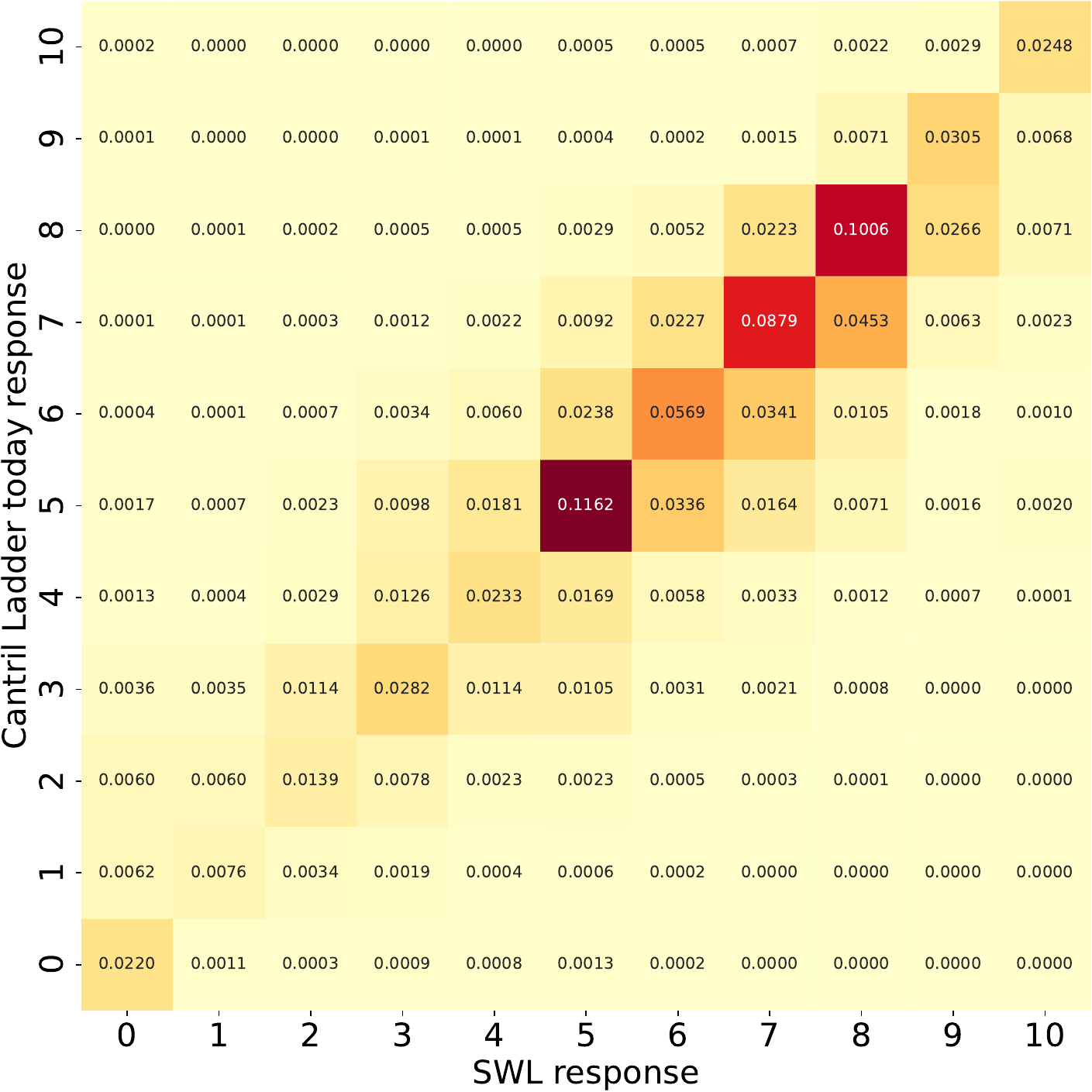} & \includegraphics[width=68mm]{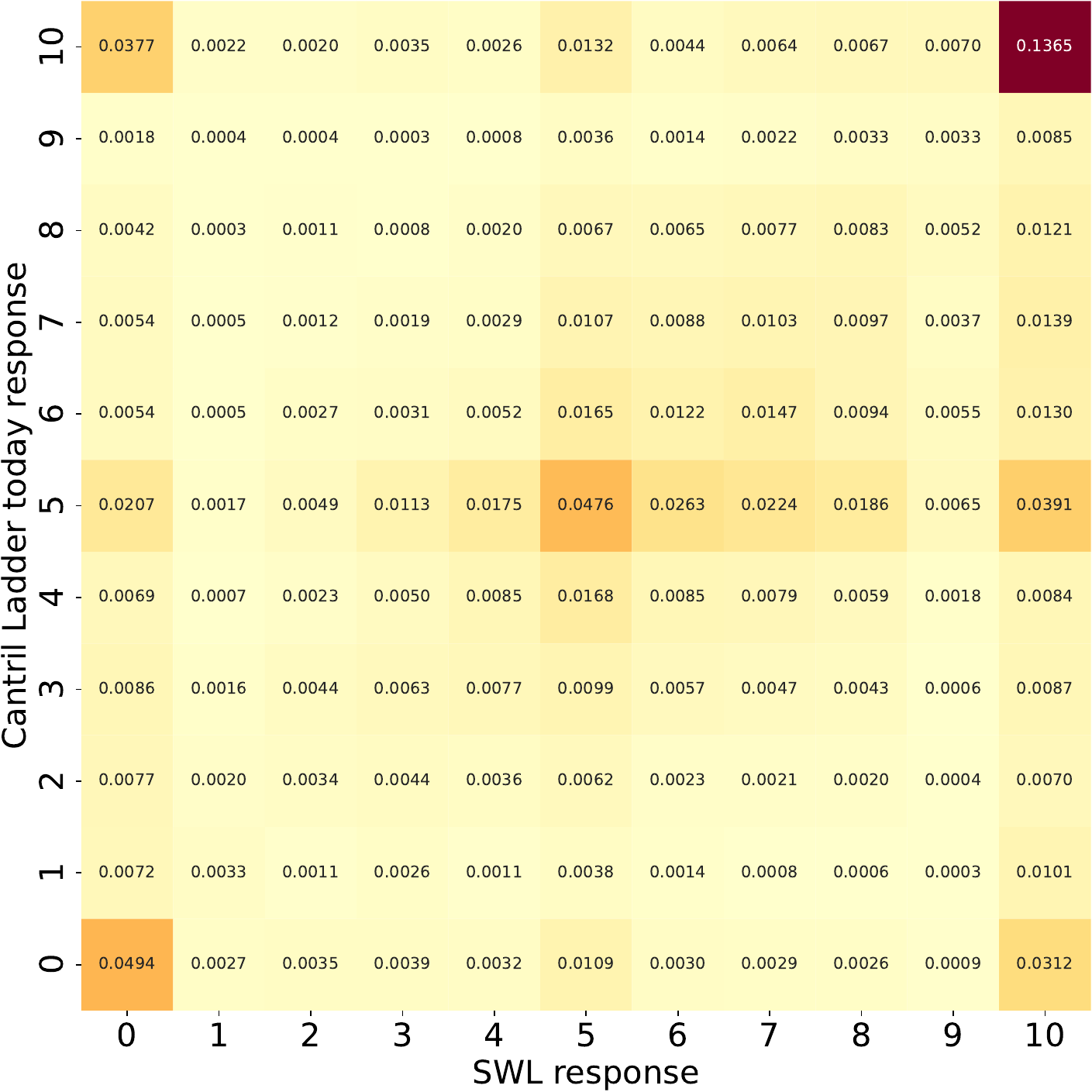} & %
\centering\rotatebox{90}{Kenya}\tabularnewline
\centering\rotatebox{90}{South Africa} & \includegraphics[width=68mm]{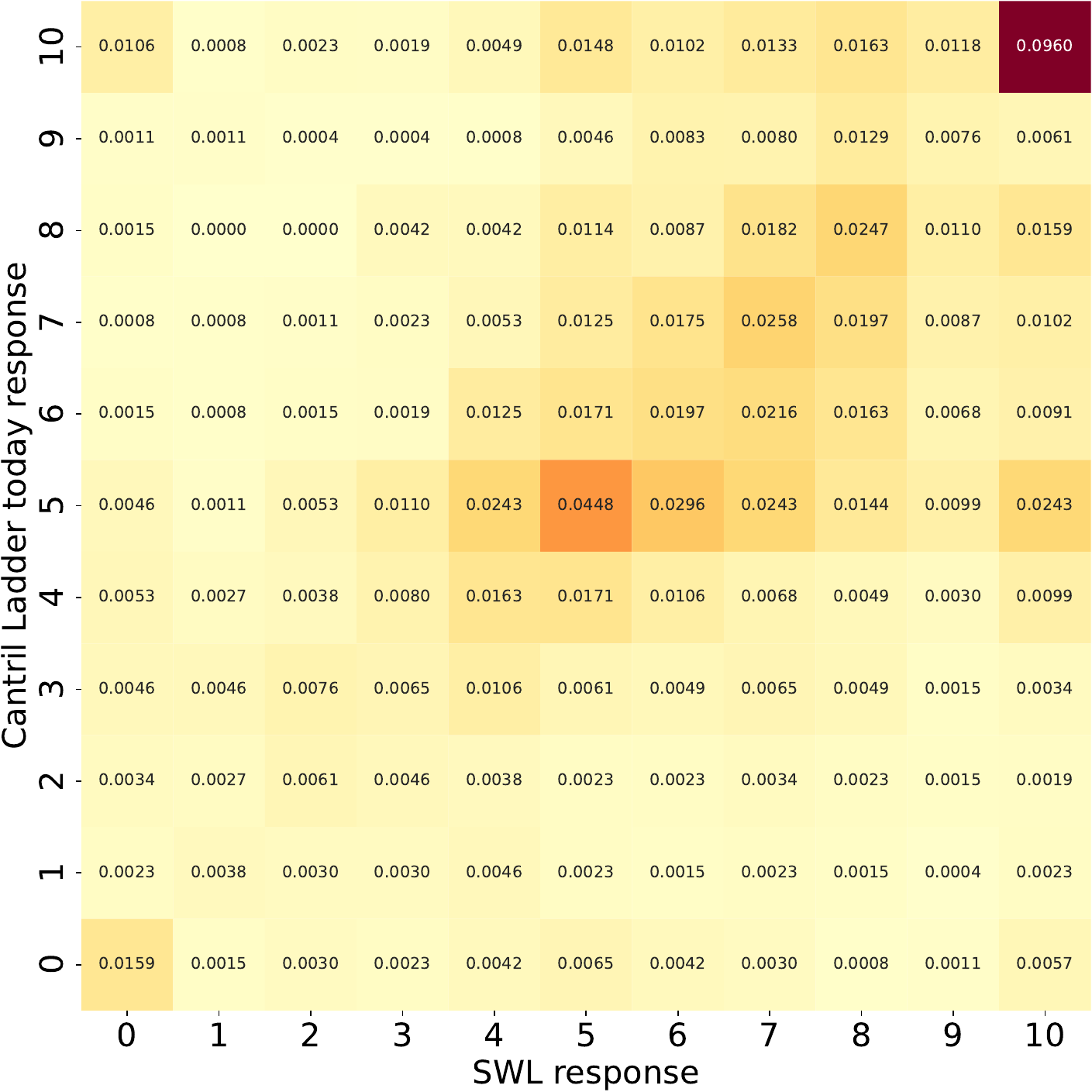} & \includegraphics[width=68mm]{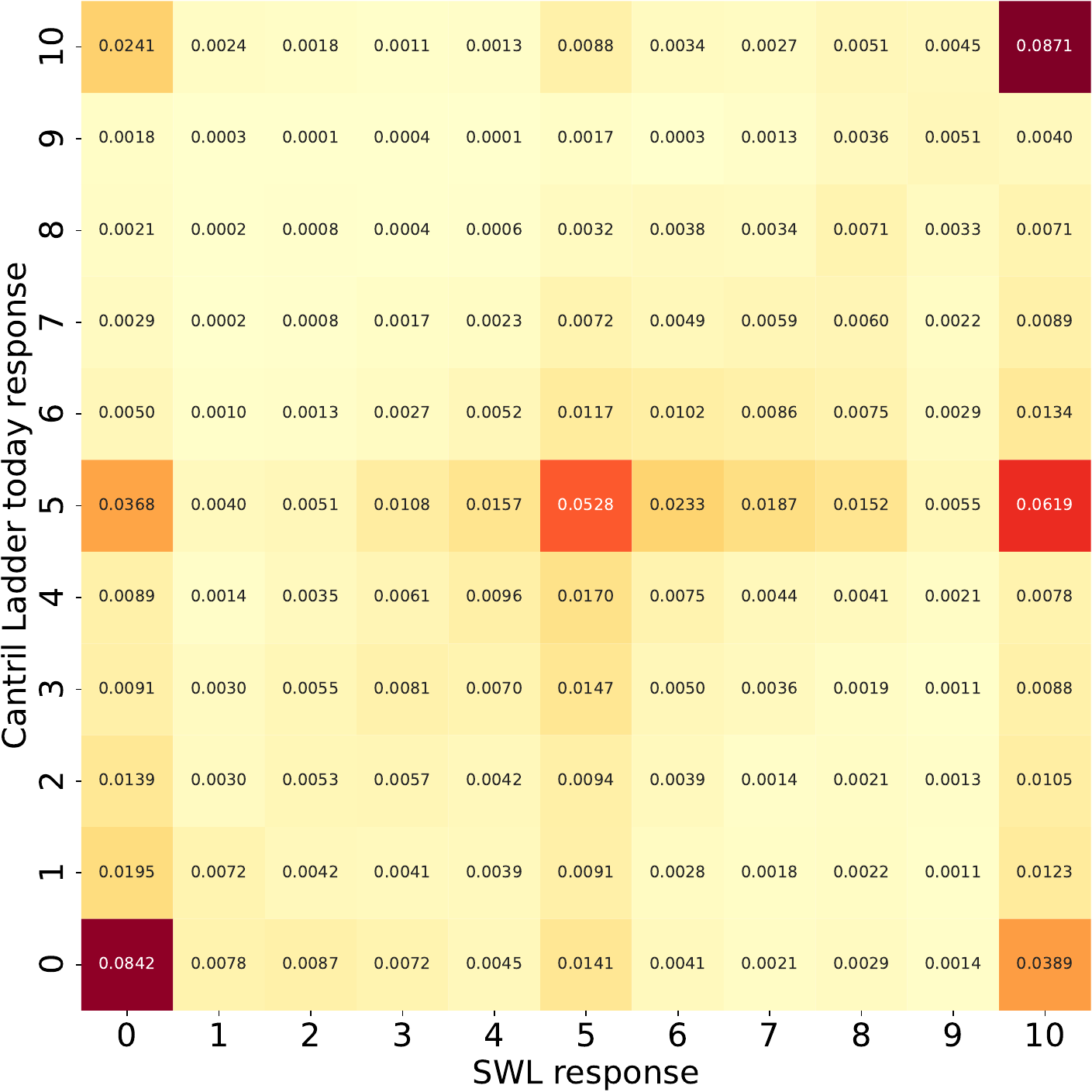} & %
\centering\rotatebox{90}{Tanzania}\tabularnewline
\end{tabular}\caption[Joint distribution of individuals' responses to life satisfaction
and Cantril ladder]{Cross-tabulations for responses to the life satisfaction question
(horizontal axis) and Cantril ladder (vertical axis) for six countries
in the GFS (2023), in which respondents answered both questions. The
fraction of total responses is noted in each cell. Within each grid
(country), darker colors correspond to higher fractions.\protect\label{fig:Joint-distributions}}

\end{figure}

The previous section, focusing on individual response distributions,
shows that differences across question wording or across surveys,
and indeed over time (\figref{distributions-grid-over-time}), all
vary by country, and that in some cases these differences are consistent
within cultural groups of countries. Because respondents in the the
GFS and early waves of the GWP answered both forms of the life evaluation
question, there is one other kind of descriptive evidence to examine
for hints to the mysteries presented so far: the joint response distribution
across the two questions. Rather than pursuing these exhaustively,
\figref{Joint-distributions} presents a small selection of six countries
from the GFS, which serves to illustrate the complexity of response
behavior.

In some countries, including the anglophone nations (not shown), Japan,
Argentina and, to a lesser extent, South Africa, respondents tend
to give the same or similar responses to the two questions. In Egypt,
by contrast, no matter what answer a respondent gave for Cantril ladder,
their most likely response for life satisfaction was ``10''. Conversely,
no matter what response they gave for life satisfaction \textemdash{}
with the exception of ``0'' \textemdash{} their most likely response
for Cantril ladder was ``5''. Fourteen percent of respondents accordingly
responded with ``5'' and ``10'' for the two questions. For those
who gave a ``10'' for life satisfaction, nearly half as many gave
a ``0'' as gave a ``10'' for Cantril ladder.

While, as mentioned earlier, one might hypothesize a combination of
some kind of offset between answers to the questions, combined with
strong FVR, to explain the behavior in Egyptian responses, the situation
is still stranger in Kenya and Tanazania. There, not only is the 
preference for 0, 5, and 10 highly dominant, but the answers from
an individual are nearly half as likely to be opposite as the same
across questions. For example, in Tanzania, among those who responded
with ``0'' for Cantril ladder, the most common response to the life
satisfaction question was also ``0'', yet nearly half as many responded
with ``10''. Among those who responded with ``5'' for Cantril
ladder, similar numbers responded with each of 0, 5, and 10 for life
satisfaction. 

\section{Model inference\protect\label{sec:Model-inference}}

Many departures from the ideal in how people respond to subjective
wellbeing questions have been identified and found to introduce noise
or even bias into responses, yet shown not to be large enough to threaten
the usefulness of these data \parencite{Krueger-Schkade-JPubE2008,Cheung-Lucas-QoLR2014-single-item-SWL-vs-SWLS-validity,Lucas-Lawless-JPSP2013-weather-SWL,Lucas-Freedman-Cornman-Emotion2018-short-term-stability-SWL,Ferrer-i-Carbonell-Frijters-EJ2004}.
As mentioned in the introduction, analysts are most interested in
the marginal effects estimated from models accounting for differences
or changes in life evaluations across large samples. A significant
difference in average responses to LS and CL need not imply a difference
in coefficients of interest estimated in an explanatory model. For
example, a detectable bias in (influence on) average life evaluation
from something like the current weather might introduce noise that
is uncorrelated with other causal influences of interest. Even if
such differences varied somehow by culture or language, they might
not threaten inference on the importance of physical security, positive
relationships with friends and family, productive employment, and
so on. Thus, the question arises: Are the anomalies identified so
far in this paper a threat to the kinds of inference needed to inform
policy?

\textcite[see columns 2 and 3 in Table 10.1]{Helliwell-Barrington-Leigh-Harris-Huang-2010}
already compared coefficients estimated using the same model and set
of respondents to explain life satisfaction or Cantril ladder from
the GWP. This was a relatively rich model, with both individual and
country-level explanatory factors, yet coefficients match quite closely
between the two outcomes variables. 

In a similar spirit, estimates for the two life evaluation questions
in the GFS are compared below. In order to assess whether LS and CL
provide the same guidance about marginal effects, a selection of relatively
objective life circumstances contained in the GFS is used as explanatory
variables. These are: log of household income; two education indicators
for at least secondary and post-secondary; gender; a quadratic of
age; two marriage status indicators; indicators for whether the respondent
lives in an urban area, is unemployed, has donated to charity in the
last month, drinks alcohol, and prays to a deity; a measure of how
many days per week the respondent exercises, and the log of the frequency
which which they attend religious services (see \tabref{GFS-descriptive-stats}
for descriptions).

Individual respondents' life evaluations, either LS or CL, are modeled
as a continuous normal variable, 

\begin{equation}
y_{i}\sim\mathcal{N}\big(\alpha_{j}+\mathbf{X}_{i}^{\top}\boldsymbol{\beta}_{j},\,\sigma\big)\label{eq:hierarchical-individual}
\end{equation}

where individual $i$ lives in country $j$, and $\mathbf{X}_{i}$
is a vector of $K$ individual characteristics. Intercepts $\alpha_{j}$
and slopes $\boldsymbol{\beta}_{j}$ vary by country (sometimes called
\emph{random effects}) but are estimated with Bayesian ``partial
pooling'' in order to achieve the most precise and reliable estimates
across countries. To accomplish this, the country-level parameters
are drawn from global (Normal) distributions:\footnote{Formally, $\boldsymbol{\beta}_{j}\equiv(\beta_{j1},\ldots,\beta_{jK})^{\top}.$}

\begin{align*}
\alpha_{j} & \sim\mathcal{N}(\mu_{\alpha},\,\sigma_{\alpha}),\\
\beta_{jk} & \sim\mathcal{N}(\mu_{\beta_{k}},\,\sigma_{\beta_{k}}),\quad k=1,\ldots,K,
\end{align*}
The partial pooling means that, when there is commonality across countries,
data from all countries can inform the estimates of the parameters
in these distributions. The ``hyperparameters'' characterizing those
distributions are in turn drawn from the following starting distributions:

\begin{align*}
\mu_{\alpha} & \sim\mathcal{N}(0,\,5),\quad\\
\sigma_{\alpha} & \sim\mathrm{HalfNormal}(0,5),\\
\mu_{\beta_{k}} & \sim\mathcal{N}(0,\,3),\\
\sigma_{\beta_{k}} & \sim\mathrm{HalfNormal}(0,3),\quad k=1,\ldots,K
\end{align*}
where the HalfNormal distribution is like the Normal distribution
$\mathcal{N}(0,\cdot)$ but 0 for values less than 0.

Lastly, the scalar variance of individual response noise (i.e., the
error term), $\sigma$, in \eqref{hierarchical-individual} is estimated
from another HalfNormal distribution as follows:

\[
\sigma\sim\mathrm{HalfNormal}(0,5)
\]
The large widths of the hyperpriors, compared with the eventual size
of estimated coefficients, start the model relatively uninformed,
relying on the large sample size of the GFS data set to achieve good
convergence.

\begin{figure}
\newcommand\compcoefswidth{.33\columnwidth}
\vspace{-24mm} %
\includegraphics[width=\compcoefswidth]{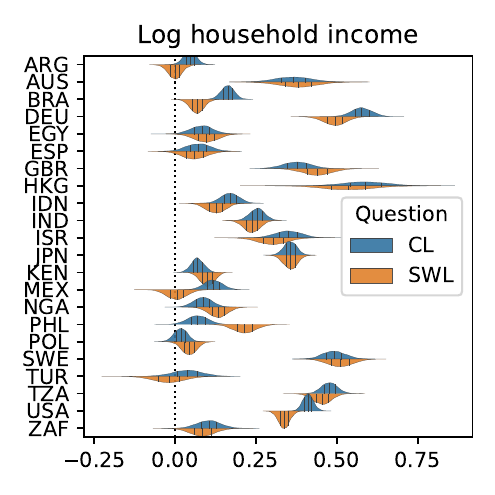}
\includegraphics[width=\compcoefswidth]{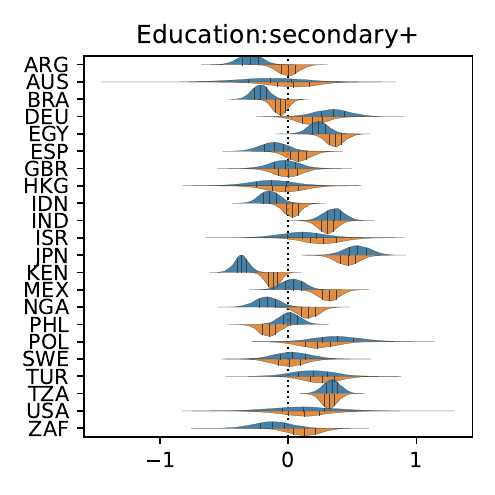}
\includegraphics[width=\compcoefswidth]{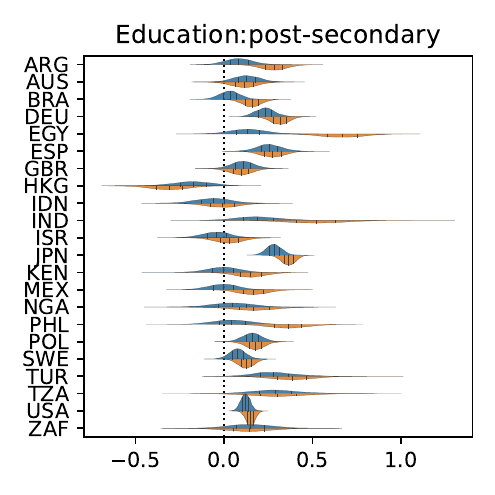}\\
\includegraphics[width=\compcoefswidth]{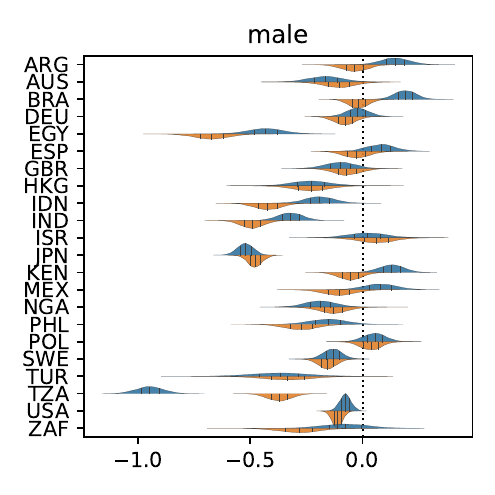}
\includegraphics[width=\compcoefswidth]{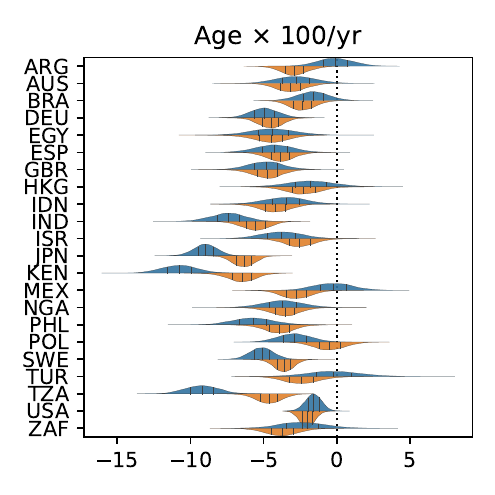}
\includegraphics[width=\compcoefswidth]{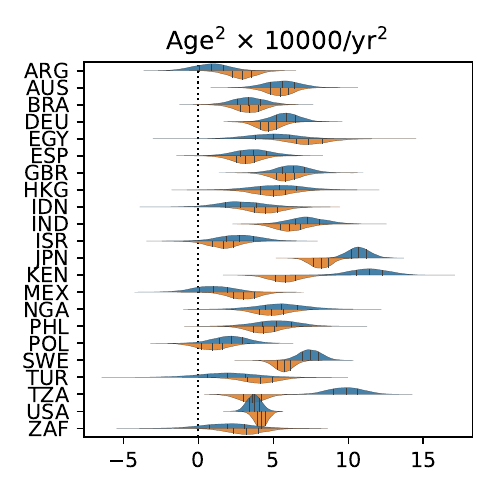}\\
\includegraphics[width=\compcoefswidth]{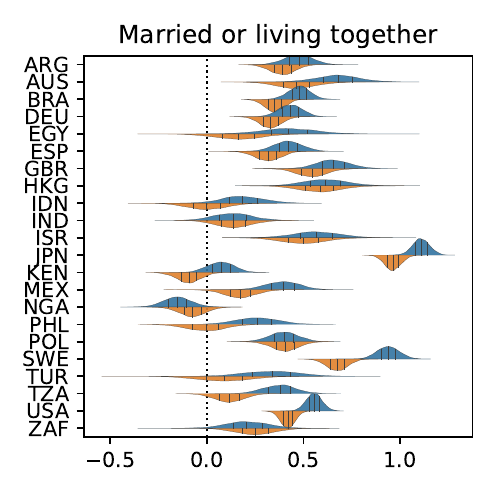}
\includegraphics[width=\compcoefswidth]{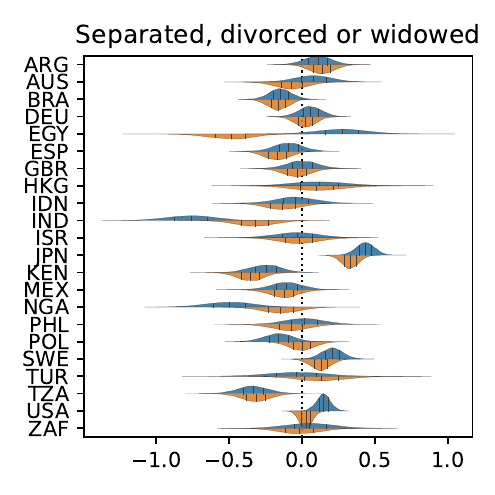}
\includegraphics[width=\compcoefswidth]{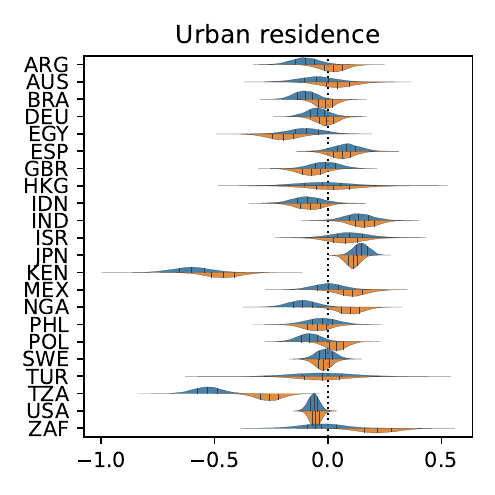}\\
\includegraphics[width=\compcoefswidth]{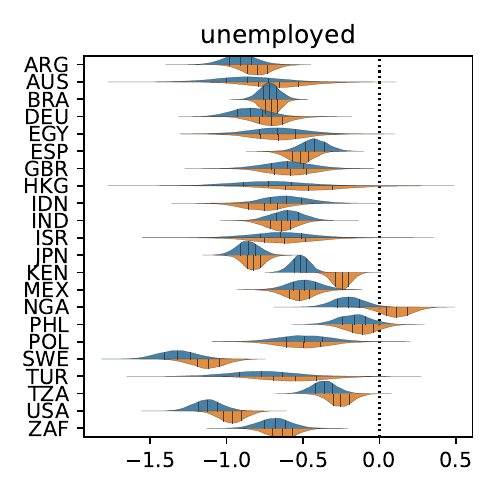}
\includegraphics[width=\compcoefswidth]{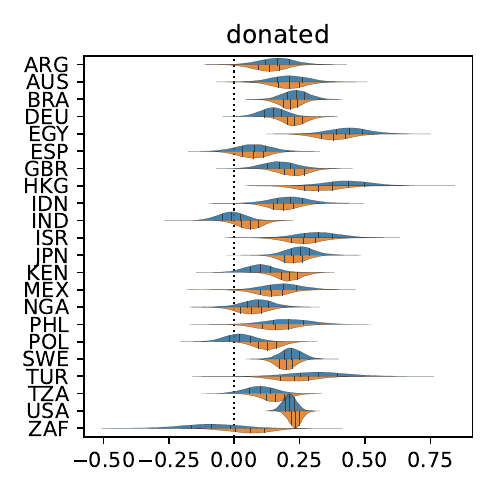}
\includegraphics[width=\compcoefswidth]{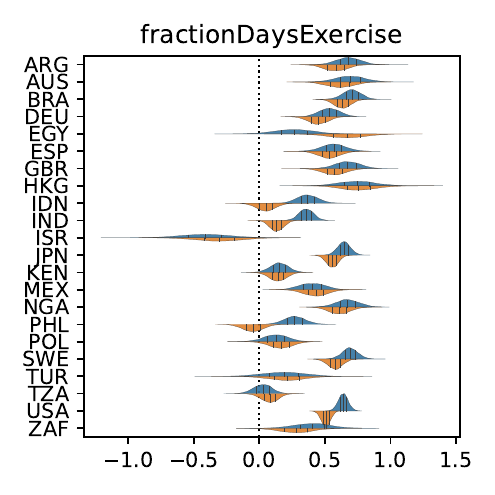}\\
\includegraphics[width=\compcoefswidth]{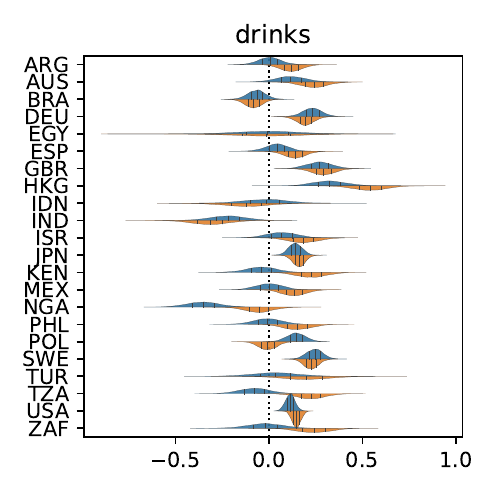}
\includegraphics[width=\compcoefswidth]{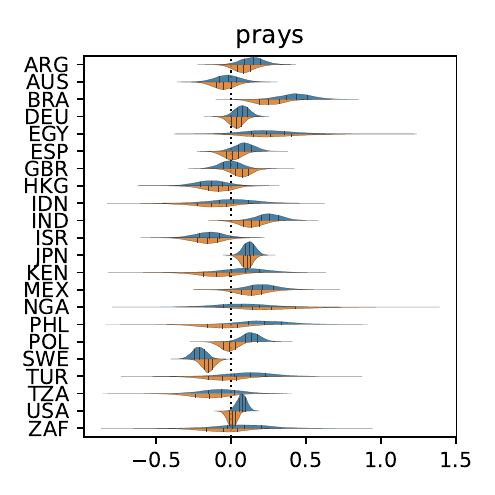}
\includegraphics[width=\compcoefswidth]{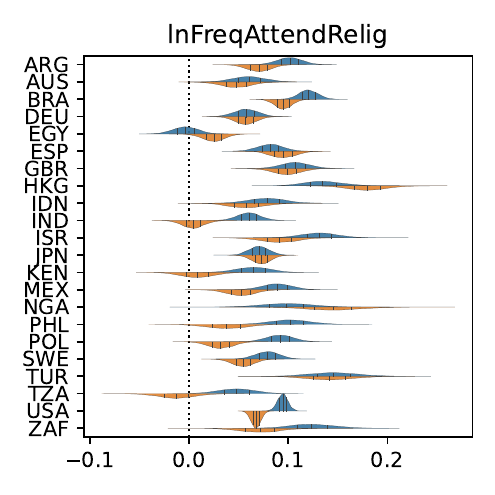}\\

\caption[Estimated coefficients using different life evaluation questions in
the GFS]{Estimates from two multivariate models, one for CL and one for LS,
both from the GFS.\protect\label{fig:compare-coefs-GFS}}

\end{figure}

\figref{compare-coefs-GFS} compares two estimates of the model, one
for Cantril ladder and one for life satisfaction. Each subplot compares
estimated country-specific coefficients for one predictor across the
two models. 

The main qualitative finding is that there is a high degree of agreement
between the two estimates. Coefficients are in many cases resolved
to vary widely across countries, yet they show strong consistency
between measures of life evaluation. Even for countries, like Egypt,
with highly divergent average responses (see Figures \ref{fig:LSGFS-vs-CLGFS-rankings}
and \ref{fig:distributions-grid-by-country}), the structure of determinants
(i.e., the coefficients in the model) is relatively consistent. There
are, of course, some strong exceptions, such as the coefficient of
gender in Tanzania. However, this case of strong statistical exclusion
is only one out of 330 pairs (22 countries $\times$ 15 predictors)
of estimated coefficients shown.

The particular specification of explanatory variables used for \figref{compare-coefs-GFS}
is driven largely by the content of the GFS questionnaire. That is,
for the present purposes the interest is more in comparing the two
questions rather than evincing policy-relevant marginal effect sizes.
Nevertheless, some comment on the magnitude of estimates is appropriate,
especially insofar as they reflect the comparability of life evaluation
questions across countries. It is important to note that the country-specific
marginal effects in \eqref{hierarchical-individual} and \figref{compare-coefs-GFS}
are estimated allowing for country-specific fixed effects $\alpha_{j}$.
Therefore, the discrepancies across countries in $\beta_{k}$ do not
necessarily represent disagreements \textemdash{} for instance, due
to cultural differences in the interpretation of life evaluation questions
\textemdash{} in the importance of aggregate levels of education,
unemployment, and so on at the national level. Rather, the coefficient
differences across countries are likely to reflect the environment
within each country, which may differ structurally.

Household income is estimated to have a typically-sized effect on
wellbeing, though it appears rather muted in some countries. By comparison
with the effect of doubling income,\footnote{A coefficient of 0.5 implies a life evaluation boost from doubling
income of $\sim0.34$ on the eleven-point scale.} the negative impact of being unemployed is enormous \textemdash{}
and relatively consistent across countries \textemdash{} as are the
positive impacts of marriage and, in most countries, regular exercise.
Social activity associated with religion also looms large, as compared
with the more individual religious activity of prayer. Education shows
a typically small or even negative impact on wellbeing after controlling
for other factors, though post-secondary education is significantly
beneficial in more developed countries. Overall, age shows a typically
U-shaped relationship to wellbeing, through a negative coefficient
on age and a positive one on the square of age.

While consistency across life evaluation questions is the dominant
pattern, there may be some subtle differences in structure evident
using the present model. The estimates suggest in some places a smaller
benefit of religious attendance on SWL as compared with CL, as well
as a smaller or more negative benefit of secondary education on CL
as compared with SWL.

\section{Discussion and Conclusion\protect\label{sec:Discussion-and-Conclusion}}

Several interconnected questions have been addressed in this study.
In more general terms than the hypotheses posed in \secref{Country-ranks},
they encompass the following three dimensions: Are the two cognitive
life evaluation questions equivalent? Is each question interpreted
the same way when asked in different surveys? Is each question interpreted
the same way in different countries? The answer to each is No, but
both the reasons for and the importance of the discrepancies remains
unresolved. Clearly, then, this paper does not solve all the problems
it raises. It is intended to lay out or spur a renewed research agenda
on intercultural and international response patterns to cognitive
life evaluation questions.

In principle, the factors affecting the answers received to life evaluation
questions include the conceptual content of a given wording of question,
the translation of the question into a local language, the framing
or context given by the rest of the questionnaire, cultural values
around the concept of a good life, cultural norms about self-expression,
and other non-cultural influences on the ability to report on the
ten or eleven point scale. This issue aligns with a recent resurgence
of interest in psychometric and econometric investigations concerning
subjective wellbeing.

Generally these efforts hypothesize some underlying functional dependence
of wellbeing on experienced circumstances, along with a reporting
function of some kind. It is hard to rationalize the joint distributions
shown in \secref{Joint-response-distributions} with a consistent
underlying distribution of wellbeing, even in light of possible focal
value rounding (FVR) behavior. It is possible that some fraction of
respondents are giving nearly random answers, in addition to simplifying
the scale. Qualitative debriefing of respondents in some of these
countries with unusual response patterns is a natural and important
approach to seek insights. 

Amid the considerable degree of reproducibility across questions within
surveys and across surveys, there may be particular anomalies plaguing
a subset of countries' responses which appear as outliers in the ranking
comparisons. The commonalities within cultural groups are certainly
suggestive of a need for revised procedures or interpretation, but
do not yet point towards a particular course of action in changing
how such data are presented. Comparisons within more homogeneous groups
would appear to be safer, yet there are cases of neighboring or culturally
related countries showing qualitatively distinct response patterns
and differences across questions and surveys.

Worldwide, the vast majority of statistical agency surveys asking
for life evaluations use the life satisfaction question, in accordance
with recommendations from the OECD and U.S. National Academies \parencite{OECD-2013-guidelines-measuring-SWB,Stone-Mackie-et-al-NationalAcademies-measuring-SWB}.
Specific criticisms of the Cantril ladder around its propensity to
evoke a focus on status and wealth rather than other life dimensions
and to elicit higher reports than life satisfaction \parencite{Nilsson-Eichstaedt-Lomas-Schwartz-Kjell-SciReports2024-Cantril-ladder-evokes-power-and-wealth}
do not go far towards explaining the cultural differences in response
differences between the two questions.

Notably, this investigation has not found evidence contradicting
the idea that marginal effects of circumstances amenable to policy
intervention can be inferred from life satisfaction regressions. While
raising problems without solutions, the evidence in this paper is
also not entirely damning even for the use of national aggregates
of the cognitive life evaluation responses in rankings. That is, no
specific fault is identified with the performance of either question,
so that it would be premature to radically change current practice
without further evidence and a deeper understanding. As the OECD prepares
its revised international guidelines for measurement of subjective
wellbeing \parencite{Mahoney-OECD2023-new-frontiers}, it is clear
that some humility, inquisitiveness, innovation, and a synergy of
investigative tools including qualitative and quantitative approaches
are called for in order to improve our confidence in measuring what
has been described as the ultimate objective in social science.

\printbibliography[heading=bibintoc]

\appendix 
\renewcommand{\thefigure}{A\arabic{figure}}
\renewcommand{\thetable}{A\arabic{table}}
\newpage ~\vfill {\centering\Huge Supplementary Material}\vfill\newpage
\setcounter{figure}{0}  \setcounter{table}{0} %

\section{Supplementary figures and tables\protect\label{sec:Appendix-tables}}

\begin{table}[h]
\begin{centering}
Life satisfaction (WVS versus GFS)
\par\end{centering}
\begin{centering}

\includegraphics{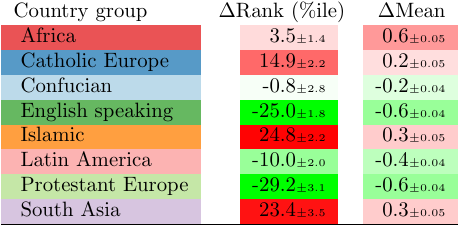}
\par\end{centering}
\caption[Average differences by cultural group for \figref{LSWVS-vs-LSGFS-rankings}]{Country group rank differences for life satisfaction responses from
WVS, as compared with GFS. Positive $\Delta$Rank means a higher ranking.
See \figref{LSWVS-vs-LSGFS-rankings} for the ranks. \protect\label{tab:table-cultgroup-mean-rankdiffs-LSWVS-vs-LSGFS}}
\end{table}

\begin{table}[h]
\begin{centering}
Life satisfaction (WVS) versus Cantril ladder (GFS)
\par\end{centering}
\begin{centering}

\includegraphics{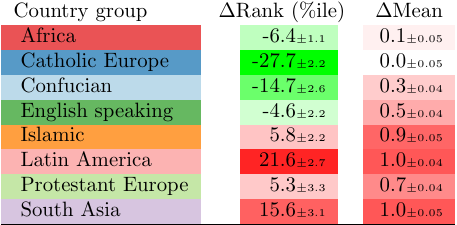}
\par\end{centering}
\caption[Average differences by cultural group for \figref{LSWVS-vs-CLGFS-rankings}]{Country group rank differences for life satisfaction (WVS) versus
Cantril ladder (GFS). Positive $\Delta$Rank means a higher ranking.
See \figref{LSWVS-vs-CLGFS-rankings} for the ranks. \protect\label{tab:table-cultgroup-mean-rankdiffs-LSWVS-vs-CLGFS}}
\end{table}

\clearpage\newpage %
\begin{landscape}%

\begin{figure}
\centering{}\includegraphics[width=0.9\linewidth]{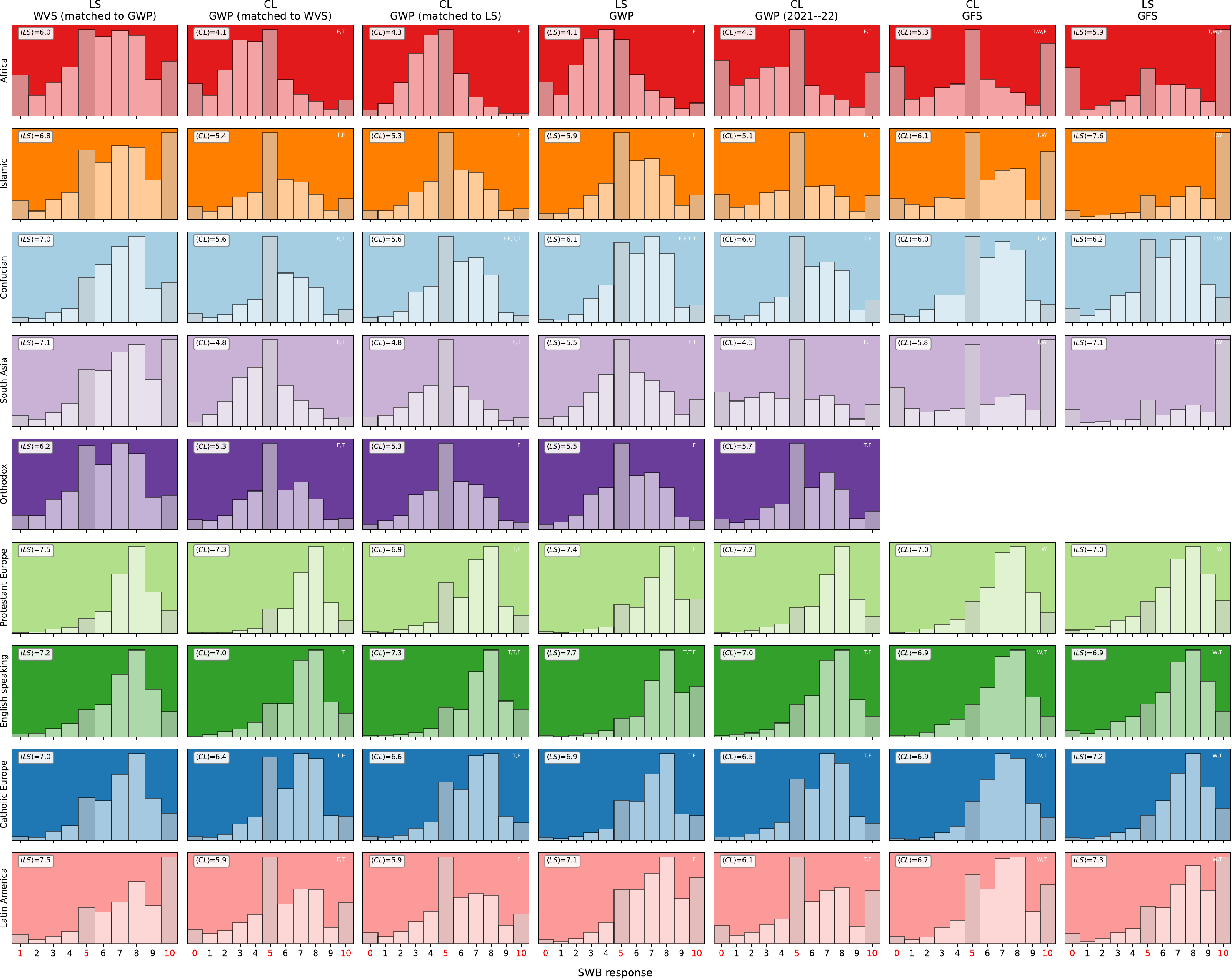}\caption{Response distributions by cultural region, for each survey and question.\protect\label{fig:distributions-grid-by-cultgroup}}
\end{figure}

\begin{figure}
\centering{}\includegraphics[width=0.9\linewidth]{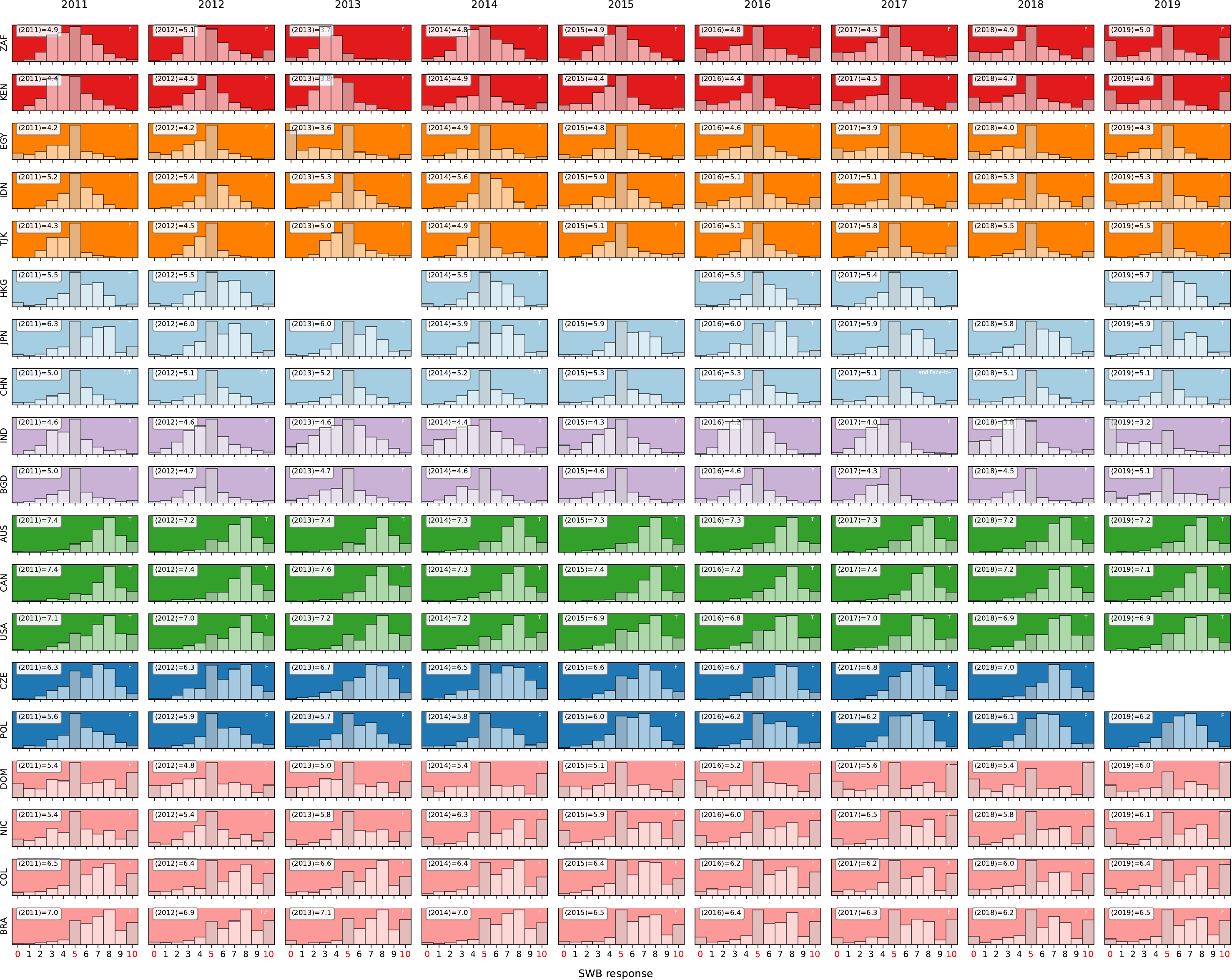}\caption[Response distributions for a selection of countries, by year.]{Response distributions for a selection of countries, by year. Generally,
patterns in the distribution are consistent across time.\protect\label{fig:distributions-grid-over-time}}
\end{figure}
\end{landscape}  %

\begin{table}
\begin{tabular}{p{3cm}rrrrrrrr}
\toprule
Variable & count & mean & std & min & 25\% & 50\% & 75\% & max \\
\midrule
LS & 202199 & 6.87 & 2.57 & 0 & 5 & 7 & 9 & 10 \\
CL & 202412 & 6.50 & 2.45 & 0 & 5 & 7 & 8 & 10 \\
lnlocalHHincome & 202898 & 1.99 & 1.26 & 0 & 1.24 & 2.01 & 2.72 & 5.30 \\ 
 &\multicolumn{8}{p{\dimexpr\textwidth-3cm-16\tabcolsep\relax}}{\footnotesize Local currencies, so magnitudes not meaningful across countries} \\
Education:secondary+ & 202711 & 0.84 & 0.37 & 0 & 1 & 1 & 1 & 1 \\
Education:post-secondary & 202711 & 0.26 & 0.44 & 0 & 0 & 0 & 1 & 1 \\
male & 202649 & 0.47 & 0.50 & 0 & 0 & 0 & 1 & 1 \\
age & 202878 & 45.83 & 17.67 & 18 & 31 & 44 & 60 & 99 \\
Married / cohabiting & 201048 & 0.62 & 0.49 & 0 & 0 & 1 & 1 & 1 \\
Separated / divorced / widowed & 201048 & 0.13 & 0.34 & 0 & 0 & 0 & 0 & 1 \\
urban & 201645 & 0.46 & 0.50 & 0 & 0 & 0 & 1 & 1 \\
prays & 202165 & 0.74 & 0.44 & 0 & 0 & 1 & 1 & 1 \\ 
 &\multicolumn{8}{p{\dimexpr\textwidth-3cm-16\tabcolsep\relax}}{\footnotesize How often do you pray or meditate? [More than once a day, about once a day, sometimes, never]} \\
donated & 202306 & 0.39 & 0.49 & 0 & 0 & 0 & 1 & 1 \\ 
 &\multicolumn{8}{p{\dimexpr\textwidth-3cm-16\tabcolsep\relax}}{\footnotesize In the past month, have you donated money to a charity? [Y / N]} \\
drinks & 199925 & 0.40 & 0.49 & 0 & 0 & 0 & 1 & 1 \\ 
 &\multicolumn{8}{p{\dimexpr\textwidth-3cm-16\tabcolsep\relax}}{\footnotesize Approximately how many full drinks of any kind of alcoholic beverage did you drink in the past seven days, if any? Please enter the number below. A full drink is a glass of wine, a can or bottle of beer, or a shot of hard liquor. [Open-ended response]} \\
lnFreqAttendRelig & 202181 & 1.07 & 2.82 & -2.30 & -2.30 & 1.10 & 3.95 & 4.64 \\ 
 &\multicolumn{8}{p{\dimexpr\textwidth-3cm-16\tabcolsep\relax}}{\footnotesize Wording: How often do you attend religious services? [More than once a week, once a week, one to three times a month, a few times a year, never]} \\
fractionDaysExercise & 199723 & 0.36 & 0.35 & 0 & 0 & 0.29 & 0.57 & 1 \\ 
 &\multicolumn{8}{p{\dimexpr\textwidth-3cm-16\tabcolsep\relax}}{\footnotesize On how many days did you exercise or engage in vigorous physical activities for 30 minutes or more in the past week? [0=0 days, 7=7 days] } \\
unemployed & 202076 & 0.08 & 0.28 & 0 & 0 & 0 & 0 & 1 \\
\bottomrule
\end{tabular}

\caption{Variables used in the GFS models of \secref{Model-inference}\protect\label{tab:GFS-descriptive-stats}}

\end{table}

\end{document}